\documentclass[conference]{IEEEtran}
\IEEEoverridecommandlockouts
\usepackage{cite}
\usepackage{amsmath,amssymb,amsfonts}
\usepackage{algorithmic}
\usepackage{graphicx}
\usepackage{textcomp}
\usepackage{xcolor}
\usepackage{multirow}
\usepackage{makecell}

\usepackage{subcaption}
\usepackage{tabularx,booktabs}
\usepackage{listings}
\usepackage{hyperref}
\usepackage{diagbox}
\usepackage{xspace}
\usepackage{pifont}
\usepackage{tikz}

\def\BibTeX{{\rm B\kern-.05em{\sc i\kern-.025em b}\kern-.08em
    T\kern-.1667em\lower.7ex\hbox{E}\kern-.125emX}}

\newcommand{\ouralg}{ShuffleV\xspace}
\newcommand{\etal}{\textit{et al.}}
\newcommand{\xmark}{\ding{55}\xspace}
\newcommand*\circled[1]{\tikz[baseline=(char.base)]{\node[shape=circle,draw,inner sep=0.1pt] (char) {#1};}}

\newcommand{\update}[1]{{#1}}

\begin{document}

\title{\ouralg: A Microarchitectural Defense Strategy against Electromagnetic Side-Channel Attacks in Microprocessors}

\author{
\IEEEauthorblockN{Nuntipat Narkthong\IEEEauthorrefmark{1}, Yukui Luo\IEEEauthorrefmark{2}, Xiaolin Xu\IEEEauthorrefmark{1}}
\IEEEauthorblockA{\IEEEauthorrefmark{1}Northeastern University, Boston, USA \IEEEauthorrefmark{2}Binghamton University, New York, USA
    \\\{narkthong.n, x.xu\}@northeastern.edu,  yluo11@binghamton.edu}
}

\maketitle

\begin{abstract}
The run-time electromagnetic (EM) emanation of microprocessors presents a side-channel that leaks the confidentiality of the applications running on them. Many recent works have demonstrated successful attacks leveraging such side-channels to extract the confidentiality of diverse applications, such as the key of cryptographic algorithms and the hyperparameter of neural network models. 
This paper proposes \textit{\ouralg}, a microarchitecture defense strategy against EM Side-Channel Attacks {(SCAs)}. 
\ouralg adopts the moving target defense (MTD) philosophy, by integrating hardware units to randomly shuffle the execution order of program instructions and optionally insert dummy instructions, to nullify the statistical observation by attackers across repetitive runs. We build \ouralg on the open-source RISC-V core and provide six design options, to suit different application scenarios. To enable rapid evaluation, we develop a \ouralg simulator that can help users to (1) simulate the performance overhead for each design option and (2) generate an execution trace to validate the randomness of execution on their workload. 
We implement \ouralg on a Xilinx PYNQ-Z2 FPGA and validate its performance with two representative victim applications against EM {SCAs}, AES encryption, and neural network inference. The experimental results demonstrate that \ouralg can provide automatic protection for these applications, without any user intervention or software modification. 
\end{abstract}


\section{Introduction}

Modern appliances and everyday objects increasingly rely on embedded microprocessors to execute software, advancing their functionality.
However, the run-time electronic characteristics of microprocessors, such as the power trace or EM emanation, present side-channels that can be leveraged to extract the confidentiality of applications. Since the demonstration of a successful side-channel attack (SCA) retrieving the encryption key of the Data Encryption Standard (DES) in \cite{kocher1999differential, ches-2004-625}, many research efforts have been devoted to attack various cryptographic algorithms (e.g., the Advanced Encryption Standard (AES) \cite{1286711, 7357115}) and develop countermeasures accordingly. 

In recent years, more edge (e.g., smart home and IoT devices) processors have become capable of running high-performance neural network models. In this way, the model inference can be performed locally with high efficiency. Although promising, these emerging systems also create a new attack surface, where an adversary can illegally extract (i.e., using side-channel analysis) the confidentiality of a neural network model. Recent works by Batina \etal \cite{Batina2019} and Takatoi \etal \cite{Takatoi2020} demonstrated successful reverse engineering attacks using side-channels to extract the neural network model architecture and weights from the microprocessors. Since producing a high-performance neural network model demands a large number of labeled data and substantial computational resources, such model architecture and parameters should be treated as the confidentiality of the model owner and well protected. Moreover, acquiring the model architecture and parameters can also aid the attacker in conducting adversarial attacks \cite{Dalvi2004} or deducing training data from the parameters \cite{carlini2021extracting}. These emerging attacks raise a new side-channel vulnerability, as the current trend is to reduce latency and power consumption while promoting {user} privacy by moving inference from the cloud to the edge processors, which makes these systems more susceptible to these side-channel-aided attacks.%

Although various countermeasures have been developed for cryptographic encryption standards \cite{Shelton2021, Singha2023} and neural networks \cite{Dubey2020, Dubey2020_ICCAD, Liu2019, luo2022nnrearch}, %
they have different limitations. For example, software-oriented solutions are easy to deploy in a microprocessor but introduce high performance overhead. On the other hand, hardware-oriented defense techniques are usually targeted customized circuit (i.e., ASIC) and are not applicable to protect general microprocessors. 
More importantly, most of these existing countermeasures are application-driven e.g., only for securing the encryption key of AES, significantly limiting their applicability.
In contrast, microarchitectural-level defense is a promising direction that can automatically safeguard any software executing on a processor without additional efforts from developers or any code modifications.

To embrace these advantages of architectural defense, this paper proposes \textit{\ouralg}, a microarchitectural defense strategy against EM {SCAs}. \ouralg adopts the moving target defense (MTD) philosophy at the architectural level. Specifically, \ouralg integrates hardware units to randomly shuffle the execution order of program instructions and randomly insert dummy instructions to prevent adversaries from using the statistical information for side-channel analysis. As a result, \ouralg can provide automatic protection against EM SCAs. 

To ensure our findings are reproducible and broadly applicable to real-world systems, we build \ouralg on the open-source RISC-V core and evaluate its performance on a Xilinx XUP PYNQ-Z2 FPGA board. %
Specifically, we extend the Ibex RISC-V core \cite{ibex}, an in-order, single-issue core with 2 pipeline stage. Being fully compatible with the original Ibex core interface-wise, \ouralg can also be used as an drop-in replacement core in the OpenTitan SoC \cite{opentitan} and several SoC in the PULP platforms \cite{PULP} like PULPissimo, PULPino, and OpenPULP. 

We make the following contributions in this work. 

\begin{itemize}

    \item We propose \ouralg, a side-channel resistant RISC-V core which integrates hardware units to randomize instruction execution order of any program to thwart EM SCAs without developer intervention or software modification.
    
    \item We implement \ouralg on FPGA and evaluate its performance with diverse workloads, including CoreMark benchmark~\cite{coremark}, AES-128 encryption \cite{AES}, and neural network inference on TensorFlow Lite Micro~\cite{tflite_micro}. To the best of our knowledge, this is the first work discussing countermeasures on a general-purpose processor that include performance and security analysis for neural network workload.

    \item %
    We perform correlation electromagnetic attacks (CEMA) with two representative victim applications, AES and neural network, to evaluate the security performance of \ouralg. For a fair comparison, we establish baseline references using the well-established open-source RISC-V core Ibex~\cite{ibex} and its security-enhanced version Secure Ibex\footnote{We refer to the Ibex core with dummy instruction insertion feature enabled as Secure Ibex in this paper.}. The experimental results demonstrate the effectiveness of \ouralg in securing these two workloads. %

    \item We implement \ouralg\footnote{\ouralg core is open-source and available at \url{https://github.com/nuntipat/ShuffleV-Demo-System}.} as an drop-in replacement to the open-source Ibex core, which enables our method to be used in many existing SoC design in the  PULP and OpenTitan project that is compatible with the Ibex core. 
      
    \item We equip \ouralg with 6 design options to suit different design objectives. To facilitate fast performance emulation, we develop a \ouralg simulator\footnote{\ouralg simulator is open-source and available at \url{https://github.com/nuntipat/ShuffleV-Simulator}.} to simulate the performance overhead and estimate the security enhancement on each workload. 
\end{itemize}

\section{Background and Related Works}

\subsection{Side-channel Attack Overview} 
Side-channel attack (or SCA) is a type of advanced attack threatening a wide range of systems, including but not limited to cryptosystems. Unlike traditional attacks that exploit the deficiency of software programs, SCA focuses on passively observe, collect, and analyze the information gleaned from the physical components during the system execution, from which the attacker can indirectly divulge information about the victim applications. The representative side-channels include the power trace \cite{kocher1999differential}, EM emanation \cite{Batina2019, Takatoi2020}, and timing information \cite{Gerlach2023}.

Correlation Electromagnetic Attack (CEMA) is a specific variant of SCA that studies the correlation between the EM emanation of a victim device and the data it processes, to extract secret information. CEMA necessitates collecting and statistically analyzing a large set of EM traces from the device under test (DUT). %
The initial step in this process is to create an appropriate EM model that reflects a certain part of the DUT, i.e., a module like a register that retains intermediate computational values reliant on the secret. Following this, a temporal correlation is identified within the EM trace at when the register updates with the intermediate computational value. The predicted EM values from the model are then cross-referenced with the actual EM traces. Given that the secret within the EM model is unknown, all potential secret values are evaluated through hypothesis testing. The accurate value is the one that yields the highest correlation between the model predictions and the actual measurements.

CEMA has been proven effective in retrieving secrets from microprocessors, such as the cryptographic key of the Advanced Encryption Standard (AES)~\cite{danial2020scniffer}, a long-term victim example in side-channel research. More recently, {EM SCAs} are also used in reverse engineering the confidentiality of neural network models from microprocessors \cite{Batina2019, Takatoi2020}, i.e., to extract the neural network architectures and weights.

\subsection{Defense against {SCAs}}

In recent years, various defenses against Side-Channel Attacks (SCAs) have been proposed. Initially driven by attacks on cryptographic algorithms, most existing defenses are designed to secure implementations of ciphers like AES by protecting sensitive information such as cryptographic keys. Following the successful application of side-channel analysis to reverse-engineer neural network models, several defenses have also emerged to protect Deep Neural Network (DNN) model architectures and weights.

Many defenses have been developed for both hardware and software implementations of cryptographic algorithms. On the software side, a notable work is Rosita \cite{Shelton2021}, a program rewriting engine that automatically protects masked AES implementations based on the input of a leakage emulator. For hardware, Bilgin \etal \cite{10.1007/978-3-319-06734-6_17, 7079468} and De Cnudde \etal \cite{10.1007/978-3-319-31271-2_16} proposed secure hardware implementations based on Threshold Implementations (TI) \cite{10.1007/11935308_38}, securing against first-order and second-order power analysis attacks, respectively. Building on this, Gross \etal \cite{HannesDOM2016} introduced Domain-Oriented Masking (DOM), a technique offering comparable security to TI with reduced chip area and fewer random bits.

For neural network workloads, Liu \etal \cite{Liu2019} proposed obfuscating memory access patterns by shuffling NN weight memory accesses and adding dummy access signals. Luo \etal \cite{luo2022nnrearch} presented more advanced scheduling obfuscation methods to protect against DNN architecture reverse engineering on FPGA-based DNN accelerators. Beyond shuffling, masking techniques have been explored by Dubey \etal \cite{Dubey2020, Dubey2020_ICCAD} to protect linear and nonlinear operations in hardware neural network implementations. Their results indicated an overhead of 3.5\% in latency and a 5.9x increase in area on a Xilinx Spartan-6 FPGA.

In addition to application-specific defenses, another line of work focuses on developing generalized defense strategies to mitigate side channels for diverse applications at the microarchitecture level, using techniques such as shuffling, code morphing, and masking. Bayrak \etal \cite{Bayrak2012} utilized a customized hardware unit to randomize the execution order of independent instruction blocks, which are either manually defined by the developer or detected during compile time using their proposed toolchain. Antognazza \etal \cite{Antognazza2021} proposed Metis, an integrated hardware module for transparent code morphing at the microarchitecture level, replacing each target instruction with several equivalent instructions. While their approach significantly reduced execution time overhead by 21-141x compared to software-based continuous code morphing, their AES-128 encryption execution time remained 1.8 to 2.35 times slower than that of an unprotected core. Gross \etal \cite{Hannes2017} employed the DOM technique to protect the register file, ALU, and data memory interface of the V-scale RISC-V processor. The evaluation results showed that their design secured the Authenticated Encryption Scheme (ASCON) \cite{ASCON} against first-order attacks, although with a 1.59x increase in LUTs and a 1.84x increase in registers on a Xilinx Spartan-6 FPGA.

\ouralg falls into this category of generalized defenses. It proposes integrating hardware units to randomize the instruction execution order of any program to thwart EM SCAs without requiring developer intervention or software modification. Our approach improves upon \cite{Bayrak2012} by eliminating the need for source code modification or recompilation with a custom compiler toolchain. Compared to \cite{Antognazza2021}, our instruction shuffling approach incurs significantly less execution time overhead because it creates execution randomness by permuting actual program instructions, rather than performing code morphing, which expands one instruction into several to obfuscate the actual computation.

\section{Threat Model, Setup, and Baseline}

\subsection{Threat Model} \label{subsec:threat_model}

\update{
This work focuses on microcontrollers and small embedded processors used in embedded and edge IoT devices, which are directly accessible to users and attackers. %
We take the EM side channel attack as case study, for the following reasons:
\begin{itemize}%
    \item \textit{Single-tenant:} These types of systems run bare-metal software or real-time operating systems (RTOS), such as \cite{FreeRTOS, Zephyr}, which contain only trusted manufacturer firmware. Therefore, we can eliminate common attacks in PC, mobile, and cloud environments based on cache, timing, branch predictors, and other microarchitecture side-channels that are launched by side loading malicious software onto the device, such as \cite{Gerlach2023, ahmadi2021side,gonzalez2019replicating}. 
    
    \item \textit{Location:} These types of system are wildly deployed on the edge and can be easily disassembled and probed. Thus, they are more vulnerable to physical SCAs.
    
    \item \textit{Non-invasive:} EM is one of the most feasible and applicable attacks in real-world scenarios, as it doesn't require physical modifications to the printed circuit board (such as removing the capacitor or cutting the power trace), unlike power side-channel attacks. It also doesn't cause irreversible damage to the target device like other invasive attacks, e.g., voltage/clock glitching, and laser injection.
\end{itemize}
}
We assume that the attacker has physical access to the device to measure the EM emanation and can send in the input and read the result of the computation from the device.

\subsection{Experimental Setup} \label{subsec:general_experimental_setup}

\begin{figure}
    \centering
    \includegraphics[width=\linewidth]{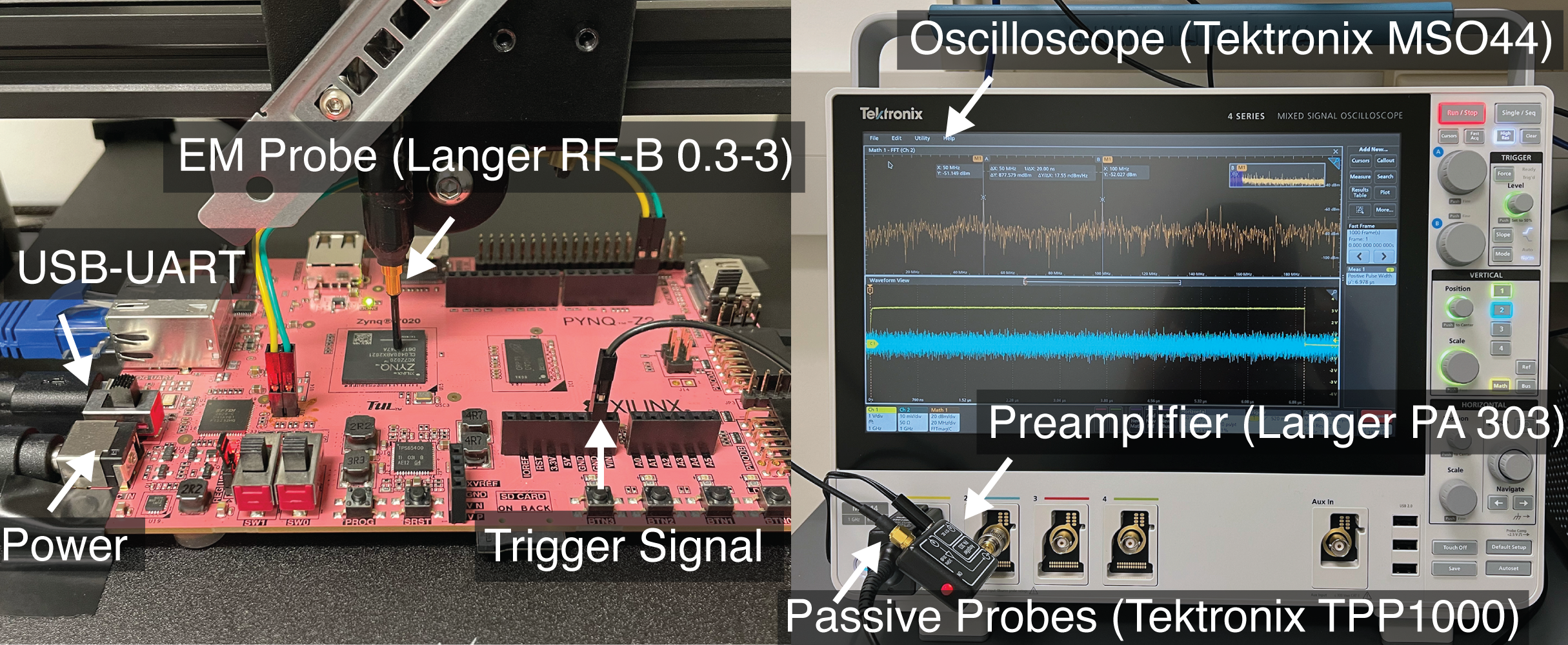}
    \caption{Our setup for performing EM measurements.%
    }
    \label{fig:ex_device}
\end{figure}

Our experimental setup is shown in Fig.~\ref{fig:ex_device}, which consists of a Xilinx XUP PYNQ-Z2~\cite{pynq-z2} FPGA board equipped with a ZYNQ XC7Z020-1CLG400C System-on-Chip (SoC). All hardware designs are synthesized using the Xilinx Vivado Design Suite 2023.1 and run at 50 MHz. We employ Tektronix MSO44 4-BW-1000 mixed signal oscilloscope~\cite{Oscilloscope}, Langer RF-B 0.3-3 H-field EM probe~\cite{prob}, and Langer PA 303 Preamplifier~\cite{Preamplifier} to collect the EM side-channel leakage from the FPGA chip. 

We consider two representative workloads: an AES-128 encryption (key = 16 Byte) and a $\sum_{i=1}^{5}in_{i}\times w_i$ (5i5w) Multiply Accumulator (MAC) operation, where $in$ represents the input data from the user, $w$ is the weight parameter, and $i$ denotes the $i^{th}$ input/weight. 
Following \cite{Batina2019, Takatoi2020}, we select the MAC operation to represent the neural network workload, as it is the core operation that formulates the convolutional and fully connected layers, the core building blocks of the neural network models.
To ensure fairness, one million random input samples are pre-generated for each workload and consistently used during EM trace collection across all core configurations.

To perform the attack, we feed each pre-generated input and collect one EM trace per input. The optimal position of the EM probe is determined by continuously running the target application and moving the probe in a grid pattern to find the location that yields the highest signal amplitude at the system's operating frequency. The length of each trace varies depending on the specific workload and the actual execution time of each run, as illustrated in Fig. \ref{figure:ibex_dummy_aes_trace_length}.
To simulate a highly capable attacker and evaluate each defense mechanism under a worst-case scenario, a trigger signal is added into the code to indicate the start and stop of computation, enabling synchronized EM measurements. The underlying idea is that if an attack proves unsuccessful under these ideal conditions, its likelihood of success in a real-world environment where misalignment and measurement noise are increased is significantly diminished.

\subsection{Baseline Characterization} \label{subsec:baseline_attack}

\begin{table*}[htbp]
\centering
\footnotesize
\caption{The number of EM traces required to extract AES-128's key on the targeted core configuration.}
\label{tab:baseline_attack_aes}
\resizebox{\linewidth}{!}{
\begin{tabular}{l|cccccccccccccccc}
\toprule
\textbf{\backslashbox{Config}{Byte\#}} & \textbf{0} & \textbf{4} & \textbf{8} & \textbf{12} & \textbf{1} & \textbf{5} & \textbf{9} & \textbf{13} & \textbf{2} & \textbf{6} & \textbf{10} & \textbf{14} & \textbf{3} & \textbf{7} & \textbf{11} & \textbf{15} \\
\midrule
Ibex \cite{ibex}  & 300  & 300    & 400    & 150    & 400    & 200    & 100    & 100    & 150    & 250    & 250    & 250    & 1050   & 100    & 450    & 450  \\
Secure Ibex \cite{ibex} & 80k & 51k & 93k & 31k & 53k & 43k & 18k & 10k & 10k & 13k & 13k & 10k & 12k & 3k & 14k & 1k \\
\bottomrule
\end{tabular}
}
\end{table*}

\begin{table}[t]
\centering
\footnotesize
\caption{The number of EM traces required to extract the 5i5w MAC weights on the targeted core configuration. \textsuperscript{*1}The attack was performed on weights 1 and 2 simultaneously.}
\label{tab:baseline_attack_mac}
\begin{tabular}{l|ccccc}
\toprule
\textbf{\backslashbox{Config}{Weight ($i$)}} & \textbf{1} & \textbf{2} & \textbf{3} & \textbf{4} & \textbf{5} \\
\midrule
Ibex \cite{ibex} & 270  & 170  & 300 & 150 & 670 \\
Secure Ibex \cite{ibex} & 37k\textsuperscript{*1} & 37k & 10k & 14k & 36k \\
\bottomrule
\end{tabular}
\end{table}

\textbf{Baseline Setup.} We use Ibex~\cite{ibex}, an open-source RISC-V core, and its enhanced version Secure Ibex~\cite{ibex} as our baseline. The Ibex core is chosen as it is designed specifically for embedded control applications and is equipped with various security features, such as register file error correcting code (ECC), cache ECC, dummy instruction insertion, and data-independent timing to defend against fault injection attacks as well as timing, power, and EM-based side channel leakages. More importantly, the Ibex core is widely adopted as a key component in the PULP \cite{PULP} and OpenTitan \cite{opentitan} platforms and is one of the most popular open-source RISC-V cores on GitHub, with over 1,100 stars. To evaluate these cores, we utilize the Ibex demo system \cite{ibex_demo_soc}, which provides debug support and some necessary peripherals in its default configuration. 

The Secure Ibex core is equipped with a dummy instruction insertion mechanism to enhance security by randomly adding one dummy instruction every few actual instructions. We enable this mechanism (by asserting the \texttt{dummy\_instr\_en} control bit in the cpuctrl register) and set the dummy instruction interval (DII) to 16 to insert dummy instructions every 0–16 real instructions. 
Four types of instruction are used as dummy instructions, each requiring a different number of CPU cycles: \texttt{ADD} (1 cycle), \texttt{AND} (1 cycle), \texttt{MULT} (2 cycles), and \texttt{DIV} (37 cycles). These varying cycle counts play a role in \textit{moving target} of the victim applications, e.g., making it more challenging to analyze side-channel traces from repetitive runs.

\textbf{Attacking Ibex.}  We first conduct CEMA attacks \update{using the Hamming weight leakage model} on the Ibex core~\cite{ibex}. We show the experimental results in the 1$^{st}$ row of Tab. \ref{tab:baseline_attack_aes} and \ref{tab:baseline_attack_mac}, from which we can observe that, the CEMA attack can easily extract the secret from the Ibex core, i.e., only several hundred traces are needed to extract the AES key and the {MAC} weights. 

\begin{figure}[]
    \centering
    \includegraphics[width=0.85\linewidth,trim={0 10 0 0},clip]{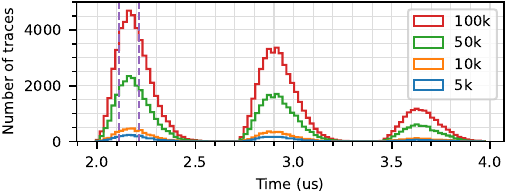}
    \includegraphics[width=0.85\linewidth]{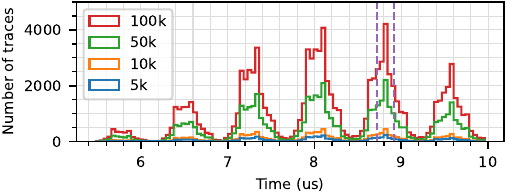}
    \caption{Histogram of the length of EM traces collected from the 5i5w MAC operation (top) and the AES-128 encryption (bottom) programs run on the Secure Ibex core. Purple dashed lines indicate the minimum and maximum lengths of traces used for the attack.
    }
    \label{figure:ibex_dummy_aes_trace_length}
\end{figure}

\textbf{Attacking Secure Ibex.} Although the Secure Ibex presents enhanced robustness against CEMA, we find it is still possible to conduct a successful attack. %
We illustrate our attacking idea in Fig. \ref{figure:ibex_dummy_aes_trace_length}, %
which shows the distribution of program execution time when capturing 5k, 10k, 50k, and 100k EM traces. We can observe that, although the insertion of dummy instructions corrupts the repetitive accumulation of side-channel features, these EM traces %
can still be clustered into a small number of groups, each following a normal distribution. Therefore, by collecting only traces of comparable length, such as the one between the two purple dashed lines, we can easily gather a substantial quantity of aligned traces. From these experiments, we found that 10k EM traces is sufficient to build an accurate distribution. Following this, an additional 100k traces of comparable length are collected and CEMA is executed to extract the secret from both AES and {MAC operation}, as shown in the 2$^{nd}$ row of Tab. \ref{tab:baseline_attack_aes} and \ref{tab:baseline_attack_mac}. 

\begin{figure}
    \centering
    \includegraphics[width=0.85\linewidth,trim={0 0 0 0},clip]{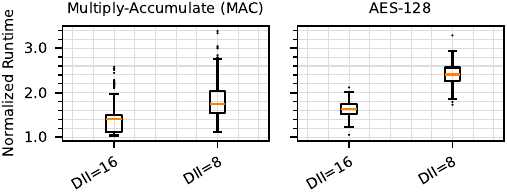}
    \caption{Normalized execution time of the Secure Ibex with dummy instruction interval (DII) = 8 and 16 compared to the Ibex core on MAC and AES-128 workload (lower is better).}
    \label{figure:dummy_overhead}
\end{figure}

\textbf{Lessons Learned.} Our experimental results \update{in Tab. \ref{tab:baseline_attack_aes} and \ref{tab:baseline_attack_mac}} demonstrate that using a dummy instruction insertion mechanism alone may not provide sufficient protection against EM {SCAs}. Although Secure Ibex can insert instructions more frequently to enhance security, e.g. every 8 or 4 program instructions, doing so incurs significant performance penalties.
Fig. \ref{figure:dummy_overhead} illustrates this overhead by comparing the execution time of the Secure Ibex core to the baseline Ibex core. For the MAC and AES-128 workloads, Secure Ibex's execution time is, on average, 37\% and 81\% higher, respectively, when DII is set to 16. This increases to 61\% and 142\% higher when DII is set to 8. Another significant drawback of Secure Ibex is its high execution time variation, which makes it unsuitable for real-time embedded systems that demand predictable timing.
To overcome these shortcomings, this paper proposes \ouralg. Our method employs a nondeterministic instruction shuffling technique to achieve superior resistance to EM SCAs, significantly reducing execution time overhead to just 13.7\% for the MAC workload and 3.1\% for the AES-128 workload, all while demonstrating much lower execution time variation.

\section{\ouralg: Design Overview} \label{sec:design_overview}

To mitigate both security and performance concerns of Secure Ibex, we propose a microarchitectural defense strategy \textit{\ouralg}, which adopts the \textit{moving-target-defense} (MTD) philosophy. 
Following our baseline characterization in Sec. \ref{subsec:baseline_attack}, we use the RISC-V ISA specifically the base integer ISA (RV32I) to introduce \ouralg. Nevertheless, the design is directly applicable to other RISC-V ISA extensions described in \cite{waterman_asanovi_2019}, e.g., compressed (RVC), multiply and divide (RVM), atomic (RVA), single/double floating point (RVFD), etc., as well as other ISAs such as ARM.

\begin{figure}
    \centering
    \includegraphics[width=\linewidth]{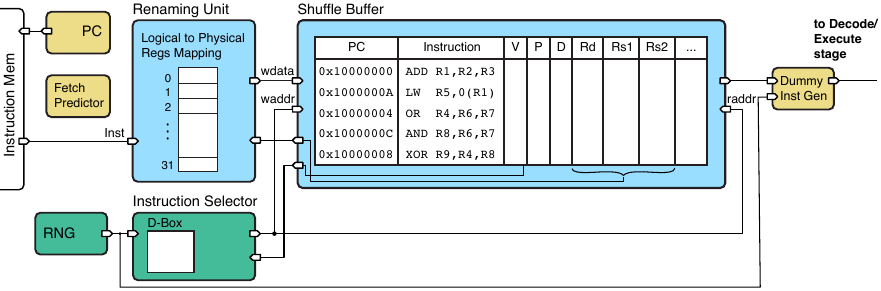}
    \caption{\ouralg's high level diagram of the instruction fetch and shuffle process.}
    \label{figure:shufflev_overall_design}
\end{figure}

The general idea of \ouralg can be described as follows: as a MTD-based defense solution, it randomizes the execution sequence of instructions. Particularly, instead of fetching and executing each instruction one-by-one, \ouralg fetches $N$ instructions to fill a hardware unit, \textit{shuffle buffer}, which stores %
the fetched instruction awaiting for execution in the next step. Each entry in the instruction buffer contains a program counter (PC) value, the machine instruction, a valid bit, $N$ dependency bits to indicate which other instructions this instruction depends on, and the index of the physical source and destination registers, as illustrated in Fig. \ref{figure:shufflev_overall_design}. Then, \ouralg randomly selects one pending instruction from the buffer to execute, and refills the buffer with new instructions retrieved from the instruction memory or an instruction cache, if available. 
Note that the shuffle buffer size is reconfigurable, i.e., can be arbitrarily chosen. While a large buffer size helps to improve execution randomness, it also incurs more hardware and execution time overhead. In our experiments, we found that a buffer size equal to 4 yields a good trade-off between security and hardware overhead (see Sec. \ref{sec:evaluation}).

\subsection{Instruction Dependency Tracking} \label{subsec:dependency_tracking}

Although the MTD philosophy is straightforward, its implementation is challenged by \textbf{how to track data dependency between instructions to maintain program correctness.} 
In an ideal scenario, there will be $N^M$ ways that a program can be executed, where $N$ is the size of the shuffle buffer and $M$ is the number of CPU cycles required to execute the program, with no performance penalty aside from the larger hardware footprint for the shuffle buffer and the auxiliary logic. 
In practice, however, not all instructions can be selected for execution in each cycle, due to data dependency between instructions. As a result, the number of unique permutations of the program instructions will be lower than $N^M$, as shown in Fig. \ref{figure:summary_pt_cycle_vs_no_ready_inst}. 
Moreover, control flow statements such as branch or jump also prevent the core from fetching new instructions, while waiting for the branch or jump to execute. Thus, the processor will need to stall for a few cycles after every branch or jump to refill the shuffle buffer, negating the overall performance of the processor, as shown in Fig. \ref{figure:shufflev_wo_speculative_fetch}. 

In RISC-V, all instructions perform computation on the register with one destination register and two source registers, except a few instructions in RVF and RVD extensions that have three source operands. 
The RISC-V's common instruction format and addressing mode make it an ideal candidate for implementing our instruction shuffling strategy.
Leveraging this feature, \ouralg checks whether the next instruction depends on any previous instruction, by comparing the next instruction source register (\texttt{rs1}, \texttt{rs2}, and \texttt{rs3}) with destination registers (\texttt{rd}) of all pending instructions. This simple strategy covers most instructions in the RISC-V ISA, except the load/store, synchronization, environment, and control status register instructions, which require special attention.

One critical optimization to maximize the ``shufflability'' of the instruction stream is to eliminate artificial dependencies, e.g., the Write-After-Write (WAW) and the Write-After-Read (WAR). These dependencies are caused by the limited number of logical registers in the ISAs, which enforces the compiler to reuse the registers during code generation. In the following example, all instructions need to be executed sequentially as there is a true data dependency between the ``\texttt{LW} and \texttt{SUB}'' instructions and the ``\texttt{AND} and \texttt{OR}'' instructions. Also, the \texttt{LW} and \texttt{AND} instructions exhibit a WAW dependency, and the \texttt{SUB} and \texttt{AND} instructions exhibit a WAR dependency. However, the \texttt{AND} and \texttt{OR} instructions are actually independent from the \texttt{LW} and \texttt{SUB} instructions thus could be executed before the first two instructions.
\begin{center}
\begin{tabular}{c}
\begin{lstlisting}
LW  R1, 4(R1)   // R1 = MEM[R1 + 4]
SUB R3, R1, R2  // R3 = R1 - R2
AND R1, R4, R5  // R1 = R4 & R5
OR  R7, R1, R8  // R7 = R1 | R8
\end{lstlisting}
\end{tabular}
\end{center}

We adopt register renaming, a technique widely used in super-scalar and out-of-order processor design, to address these artificial dependencies (WAW and WAR). The idea is to have more physical registers in the hardware than the number of registers in the ISAs (i.e., logical registers) and reserve new physical registers to be used as destination registers for every instruction that writes to the register file. The current mapping between logical and physical registers is maintained in the renaming unit. The corresponding physical registers for all source and destination registers of all pending instructions are recorded in the corresponding fields in the shuffle buffer. These data are needed to determine unused physical registers, i.e., registers that are not referred to in the logical-to-physical mapping table or in any valid entry in the shuffle buffer. 

We use the following example to illustrate the effectiveness of register renaming, where \texttt{Xn} denotes physical registers and \texttt{Rn} denotes logical registers. Assigning \texttt{R1} to \texttt{X6} in the \texttt{LW} and \texttt{SUB} instructions and to \texttt{X8} in the \texttt{AND} and \texttt{OR} instructions yields greater flexibility, regarding the execution order of these instructions. The \texttt{AND} instruction can be moved prior to the \texttt{LW} or \texttt{SUB} instruction, or alternatively, the \texttt{SUB} instruction can be moved subsequent to the \texttt{AND} or \texttt{OR} instruction.

\begin{center}
\begin{tabular}{c}
\begin{lstlisting}
LW  X6, 4(X1)   // Initial allocation 
SUB X7, X6, X2  // X1=R1 X2=R2 X3=R4
AND X8, X3, X4  // X4=R5 X5=R8
OR  X9, X8, X5  
\end{lstlisting}
\label{list:offset_chalenge}
\end{tabular}
\end{center}

Next, we explore other instructions that require special consideration, apart from verifying source/destination registers. 

\textbf{Load and Store\footnote{To support memory-mapped I/O, load and store optimization can be disabled during core configuration or temporarily via the Control and Status Registers (CSR) as described in Sec. \ref{sec:sv_implementation}}.} The load and store instructions can have an implicit dependency on each other, as the target address is calculated by the value of the register plus an integer offset, which is not known at the instruction fetching stage, until all prior instructions that write to that specific register have been executed. The simplest method is to assume that every load and store depends on all prior loads and stores, which is trivial to implement with low hardware overhead but could incur large run-time overhead. 
One exception is when there are multiple load instructions in the shuffle buffer without any store instruction. In this case, we can allow these load instructions to be executed in any particular order, given that they don't depend on other instructions in the shuffle buffer. 

More challenging, a dependency can exist between ``load and store'' and ``store and store'' instructions. For instance, the \texttt{SW} instruction in the following code example can't be executed before the first \texttt{LW} instruction, as \texttt{R1} will contain an incorrect value if \texttt{R2+4} equals \texttt{R4}. However, the last \texttt{LW} instruction can be executed anytime since we can ensure that the \texttt{SW} instruction will write to a different memory location (\texttt{R4} $\neq$ \texttt{R4+4}). Therefore, we can conclude that a dependency exists between a pair of load and/or store instructions when their base register is different, as the register value is not known. When the base register is identical, the offset must be compared depending on whether the instruction worked on a byte (\texttt{LB}, \texttt{SB}), a half-word (\texttt{LH}, \texttt{SH}), or a word (\texttt{LW}, \texttt{SW}).
\begin{center}
\begin{tabular}{c}
\begin{lstlisting}
LW  R1, 4(R2)  // R1 = Mem[R2+4] 
SW  R3, 0(R4)  // Mem[R4] = R3
LW  R5, 4(R4)  // R5 = Mem[R4+4]
\end{lstlisting}
\end{tabular}
\end{center}

\textbf{Other Instructions.} We also need to ensure that synchronization instructions (\texttt{FENCE(.I)}), environment instructions (\texttt{ECALL} and \texttt{EBREAK}) and control status register instructions (\texttt{CSR}) are not shuffled around, since they can affect run-time behavior and debugging. To achieve this goal, we set a dependency-bit to cascade these instructions, i.e., these instructions will depend on all prior instructions, and their subsequent instructions will depend on them.

\subsection{Instruction Selection}

\begin{figure}
    \centering
    \includegraphics[width=0.8\linewidth]{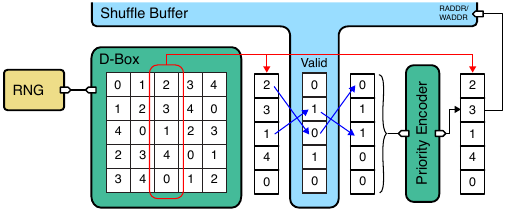}
    \caption{Instruction selection scheme in \ouralg.}
    \label{figure:shufflev_instruction_selection}
\end{figure}

After filling the instruction buffer, the next step is to pick one ready instruction from the instruction buffer to execute. 
Although the task appears simple initially, creating random numbers within a certain range and excluding those that correspond to indexes of invalid instructions is not trivial.
To solve this problem without introducing extra overhead, \ouralg employs a rule: it selects the closest ready instruction from the random index. 
For example, if the random number is 1, we will choose the first valid instruction at index 1, 2 (+1), 0 (-1), 3 (+2), and 4 (-2) in order. Fig. \ref{figure:shufflev_instruction_selection} describes how this logic can be implemented in hardware in three steps. 
(1) We pre-compute the index sort by the distance for each possible random number (0 to $N-1$) and store them in a table called D-Table. 
(2) We load one column of the D-Table depending on the generated random number and get the valid bit from the corresponding index in the shuffle buffer, as illustrated by the blue arrows in Fig. \ref{figure:shufflev_instruction_selection}.
(3) The priority encoder can be used to find the index of the first '1', which will be used to index a row in the selected D-box column to get the instruction index. 

\subsection{Performance Optimization} \label{subsec:performance_opt}

\begin{figure}
    \centering
    \includegraphics[width=0.9\linewidth]{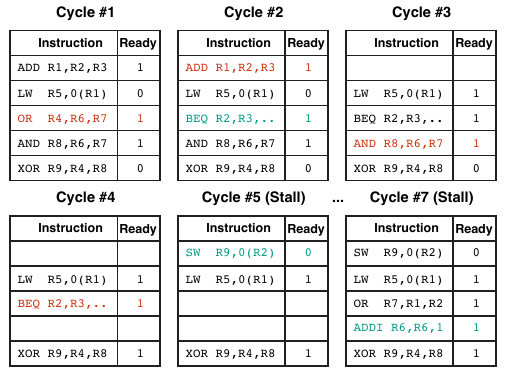}
    \caption{State of the shuffle buffer during execution without speculative fetch. Instructions highlighted in red are to be removed and executed. Instructions highlighted in green are newly fetched. The ready bit indicates that the entry is valid and is not depend on any other instruction.}
    \label{figure:shufflev_wo_speculative_fetch}
\end{figure}

As a MTD-based defense, the security performance of \ouralg lies in its shuffling capability, i.e., the randomness of instruction execution. To maximize this, %
the instruction buffer must be kept full at every cycle that the processor selects an instruction for execution. 
One exception is in the simplest design without any speculative fetch, where we temporarily allow the instruction buffer to be partially filled when waiting for the control flow instruction (branch or jump) to be executed (see cycle \#3-4 in Fig. \ref{figure:shufflev_wo_speculative_fetch}). The drawback is that it requires the core to stall for 0-$N$ cycles to refill the instruction buffer after executing each control flow instruction (see cycle \#5-7 in Fig. \ref{figure:shufflev_wo_speculative_fetch}). The number of stall cycles depends on how long the control flow instruction stays in the shuffle buffer, as the core won't be able to fetch the next instruction while waiting for this control flow instruction to be executed.

\begin{figure}
    \centering
    \includegraphics[width=0.9\linewidth]{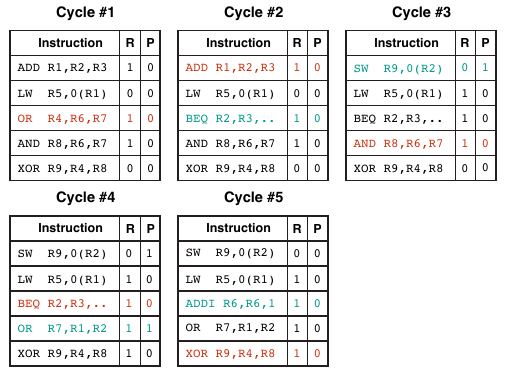}
    \caption{State of the shuffle buffer during execution with speculative fetch enabled (R=``Ready'' and P=``Prefetch''). Instructions highlighted in red are to be removed and executed. Instructions highlighted in green are newly fetched.}
    \label{figure:shufflev_w_speculative_fetch}
\end{figure}

To further improve the performance of \ouralg, we propose the following three approaches to reduce the stall cycle after each control flow instruction:

\circled{1} \textit{Speculative fetch}. One effective approach to eliminating stalls after each control flow instruction is to perform speculative fetch, by implementing branch prediction and/or return address stack. A simple implementation is to perform speculative fetch and mark the entry as prefetch, as shown in Fig. \ref{figure:shufflev_w_speculative_fetch}. When the branch or jump is executed, we can either de-assert the prefetch flag to turn the entry into a valid entry and continue execution (see cycle \#5 in Fig. \ref{figure:shufflev_w_speculative_fetch}), or clear all prefetch entries and stall to refill the buffer, depending on whether we predict the target address correctly or not.

\circled{2} \textit{Speculative execution}. Following the first approach, we can execute all prefetch instructions as if they were valid entries. To achieve this, a checkpoint is created in the same cycle that the branch or jump instruction is added to the shuffle buffer. It stores the current state of the shuffle buffer, the current physical register value, and the current mapping from logical to physical register. If the branch or jump target address is correctly predicted, the checkpoint is discarded and the execution can proceed without any overhead. However, if the target address is incorrectly predicted, we need to restore from the checkpoint and fetch the correct instruction to replace the branch or jump instruction. This approach maximizes instruction sequence randomness in exchange for higher hardware resources for the checkpoint logic.

\begin{figure}
    \centering
    \includegraphics[width=0.9\linewidth]{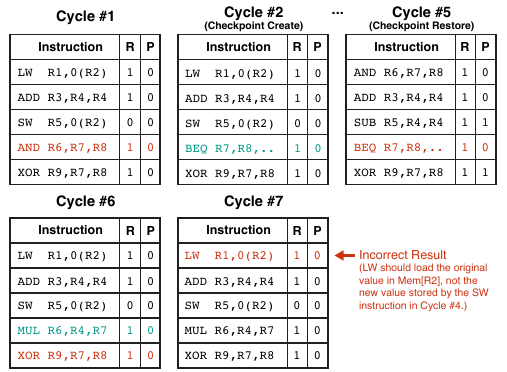}
    \caption{An example demonstrating the problem with executing dependent load and store instructions during the checkpoint period. \texttt{LW} and \texttt{SW} are executed in cycles 3 and 4, respectively.}
    \label{figure:shufflev_checkpoint_wrong_execution}
\end{figure}

Before creating the checkpoint, we must ensure that there are no dependent load and store instructions in the shuffle buffer, as the load instruction may retrieve an incorrect value from memory, if the subsequent store instruction was already executed before the rollback to the checkpoint. We illustrate this issue in Fig. \ref{figure:shufflev_checkpoint_wrong_execution}, where the checkpoint is created in cycle \#2, and the \texttt{LW} and \texttt{SW} instructions are executed in cycles \#3 and \#4, respectively. Then, in cycle \#5, the branch instruction is executed, resulting in a rollback of the core due to a branch mis-prediction. Later in cycle \#7, the \texttt{LW} instruction executes again and returns an incorrect value, as the succeeding store has already committed to memory. 

Also, during execution with an active checkpoint, it is crucial to prevent executing instructions that cannot be rolled back, such as \texttt{syscalls} that writes to I/O, etc. This can be achieved by setting the dependency bit of the relevant instructions to the branch or jump instruction. Alternately, we could adopt the load store queue and re-order buffer \cite{10.1145/285930.285988}, a hardware unit commonly used in out-of-order processor design, to commit the result of executed instructions in order and enable revert to previous checkpoints. However, doing so restricts the number of CPU cycles an instruction can be shuffled from its initial position. Additionally, committing the results sequentially may increase side-channel leakage. %

\circled{3} \textit{Modified instruction selection algorithm}. Beside introducing additional hardware for speculative fetch or execution, we can amend the instruction selection module to select the pending branch or jump instruction as soon as it is ready. %
Therefore, reducing the amount of entry in the instruction buffer that need to be refilled. This method incurs low hardware overhead, but might reduce the randomness of the instruction sequence. Practically, this approach can be combined with the first and second approaches to shorten the speculative fetch and execution duration, thus reducing the rollback penalty when the prediction is incorrect.

\subsection{Additional security feature} \label{subsec:additional_security}
To further enhance the execution randomness, \ouralg also supports inserting dummy instructions by randomly selecting one ALU operation and sending the dummy instruction to the decode/execution stage, for every 0-4, 0-8, or 0-16 instructions. The main difference between \ouralg and the Secure Ibex \cite{ibex} is the choice of dummy instructions, \ouralg uses \texttt{ADD}, \texttt{AND}, \texttt{MUL}, and \texttt{MULH} instead of \texttt{ADD}, \texttt{AND}, \texttt{MUL}, and \texttt{DIV}. The \texttt{DIV} instruction stalls the processor for 37 cycles and is key to the dummy instruction insertion in the Secure Ibex core (which we successfully attacked in Sec. \ref{subsec:baseline_attack}). Since \ouralg only utilizes dummy as an additional defensive measure, it can avoid using \texttt{DIV} instruction to significantly reduce overhead as shown in Fig. \ref{figure:overhead_shufflev_vs_others}.

\section{\ouralg: Implementation} \label{sec:sv_implementation}

\subsection{Design Options} \label{subsec:summary_shufflev_configuration}

We develop \ouralg with the following design options, to suit different application scenarios, as well as providing trade-off between security and performance overhead. %

\begin{itemize}
  
    \item \textit{Optimized memory} (M): this option allows dependencies between load and store to be determined using the logic in Sec. \ref{subsec:dependency_tracking}. Otherwise, the core assumes that all load/store instructions depend on all prior load/store instructions.
    \item \textit{Decode jump instruction} (J): this option avoids stalling when encountering the \texttt{JAL} (jump and link) instruction by adding logic to calculate the jump target immediately.
    \item \textit{Branch prediction} (B): this option uses branch prediction to reduce stall cycles. We allow only a single branch to be predicted and one checkpoint to be created at a time, to reduce the hardware overhead.
    \item \textit{Return address stack} (R): this option enables the return address stack to avoid stalling when encountering \texttt{JALR} instruction.
    
    \item \textit{Multiple checkpoint} (C): this option allows predicting multiple branches and creating multiple checkpoints. It must be specified in combination with the B option.
    \item \textit{Shortcut branch/jump evaluation} (F): this option forces the instruction selector to select the pending branch or jump instruction as soon as it is ready.
\end{itemize}

\begin{figure}[]
    \centering
    \includegraphics[width=\linewidth]{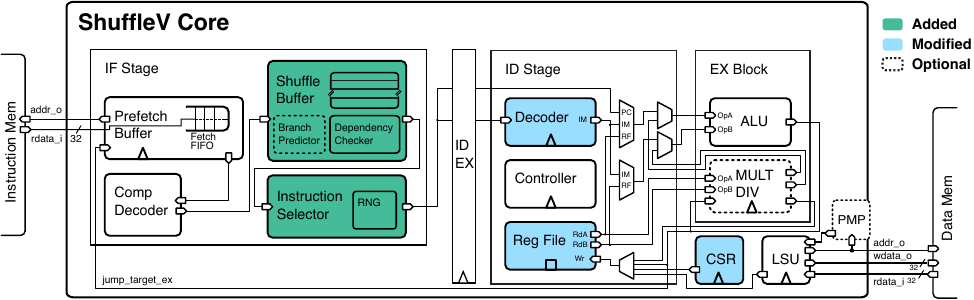}
    \caption{\ouralg core block diagram, with variations from the original Ibex highlighted. Green components are newly added. Blue components have been modified. Components with dash border are optional.}
    \label{figure:shufflev_block_diagram}
\end{figure}

\begin{figure*}
    \centering
    \includegraphics[width=\linewidth,trim={0 0 0 0},clip]{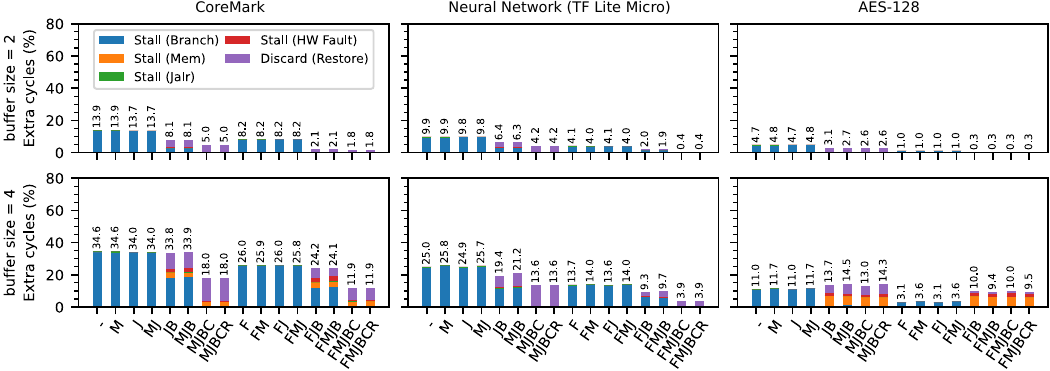}
    \caption{Performance overhead of different configuration of \ouralg on the CoreMark benchmark (left), neural network inference on TensorFlow Lite Micro (middle) and AES-128 encryption (right) with shuffle buffer size = 2 and 4}
    \label{figure:overall_pt_cycle_overhead}
\end{figure*}

\subsection{\ouralg Simulator}

To evaluate the performance and security of \ouralg, we develop \ouralg simulator based on libriscv \cite{libriscv}, an open-source RISC-V userspace emulator library. \ouralg simulator supports executing any RISC-V binary on all \ouralg configurations described in Sec. \ref{subsec:summary_shufflev_configuration}. It provides detailed execution traces, including internal state of the shuffle logic and reason for each stall, allowing a user to validate the execution randomness and select the most appropriate \ouralg configuration based on expected workload and performance requirement.

\subsection{Hardware Implementation} \label{subsec:hw_implementation}
To streamline \ouralg's integration into existing designs, we implemented it on the open-source and popular Ibex RISC-V core \cite{ibex}, an in-order, single-issue core with two pipeline stages. Ibex fully supports the base integer instruction set (RV32I) and can be configured for compressed (RV32C), multiplication and division (RV32M), and bit manipulation (RV32B) extensions. Being interface-compatible with Ibex, \ouralg can serve as a drop-in replacement in the OpenTitan SoC \cite{opentitan} and several PULP platform SoCs \cite{PULP}.

We chose to extend a simpler in-order core like Ibex over a more complex superscalar out-of-order core for two main reasons. First, this approach allows us to demonstrate the generalizability of our method on a simple, less-expensive core. Second, a more complex architecture would introduce unnecessary components. Despite our modifications, the resulting core is 3.5x smaller than state-of-the-art small out-of-order RISC-V cores, such as \cite{8977924}. A detailed comparison of the resource utilization can be found in Appendix \ref{appendix_fpga_resource}.

Fig. \ref{figure:shufflev_block_diagram} highlights the modifications made to the Ibex core. First, the shuffle buffer is introduced alongside the dependency tracking logic, the instruction selection logic, and the random number generator (RNG). The Comp Decoder is moved in front of the shuffle buffer, as our shuffling logic and the ID/EX stage work with instructions in the uncompressed form. Second, we expand the number of registers in the register file to support register renaming and modify the instruction decoder and corresponding data paths in the ID stage to refer to the (renamed) physical registers instead of the logic register specified in the machine code. Third, we add a configuration bit in the Control and Status Registers (CSR) to allow the software developer to enable/disable the protection to reduce the performance overhead. Finally, we adopt the systemc\_rng pseudo-random number generator from OpenCores \cite{opencoresOverviewSystemCVerilog}, which combines an LFSR with a CASR based on \cite{Tkacik2002}. Note that the design and evaluation of secure RNGs suitable for FPGA and/or ASIC implementation is out of the scope of this work.

Since our proposed modifications focus on the front-end (i.e., the predict and instruction fetch stages), they are compatible with any single-cycle and pipeline RISC-V core. The performance and security analysis of the implementation on other RISC-V cores are left as future work.

\section{\ouralg: Evaluation} \label{sec:evaluation}

\subsection{Performance Evaluation} 

\textbf{Evaluation Benchmarks.} We select three workloads to evaluate \ouralg, including the standard CPU benchmark (CoreMarks), neural network inference (TF Lite Micro library), and cryptography encryption (AES-128). To demonstrate compatibility with diverse neural network architectures, we employ the ensemble model (Fig. \ref{figure:tf_ensemble_model_arch} in Appendix \ref{appendix_model_arch}), \update{a small network that} consists of all widely used layers and activation function, such as fully connected, convolution, depth-wise separable convolution, batch normalization, max pooling, relu, and softmax layers.

\textbf{Execution Time.} We show the overhead of \ouralg in Fig. \ref{figure:overall_pt_cycle_overhead}, which is quantified in percent of extra (stall) cycles to complete the benchmark program on different configurations of \ouralg with different buffer sizes. These extra cycles result from the shuffle buffer needing to be refilled after certain events. The causes of these stalls are as follows:

\begin{itemize}
    \item Branch: it occurs when the core pauses fetching while waiting for the branch instruction to be selected for execution (only occurs when branch predictor is not enabled or when multiple checkpoint are not allowed).
    \item Mem: it occurs when the core pauses fetching while waiting for dependent load and/or store instructions to be selected for execution before the checkpoint can be created (only occurs when branch predictor is enabled).
    \item Jalr: it occurs when the core pauses fetching while waiting for the \texttt{JALR} instruction to be selected for execution.
    \item HW Fault: this stall is caused by an exception from a speculated instruction. In this case, the core pauses fetching and reverts to the checkpoint. 
    \item Discard: this stall is caused by a misprediction of the branch predictor or the return address stack.
\end{itemize}

Compared to the baseline unsecured core (Ibex), on the CoreMark benchmark, the execution time overhead ranges from 1.78\% to 13.95\% and 11.91\% to 34.6\% when the shuffle buffer size is equal to 2 and 4, respectively. On the neural network workload, the execution time overhead ranges from 0.39\% to 9.91\% and 3.88\% to 25.83\% when the shuffle buffer size is 2 and 4, respectively. On the AES encryption workload, the execution time overhead ranges from 0.26\% to 4.85\% and 3.1\% to 14.45\% when the shuffle buffer size is equal to 2 and 4, respectively. The overhead when buffer size is equal to 8 is given in Fig. \ref{figure:overall_pt_cycle_overhead_bs8} in Appendix \ref{ap:performance_overhead}. 
From our experiment, we found that a buffer size of 4 is a optimal compromise between performance and side-channel security (see Tab. \ref{tab:aes_cema_summary_num_trace} and \ref{tab:dot_product_cema_summary_num_trace}).

\begin{figure}
    \centering
    \includegraphics[width=0.9\linewidth]{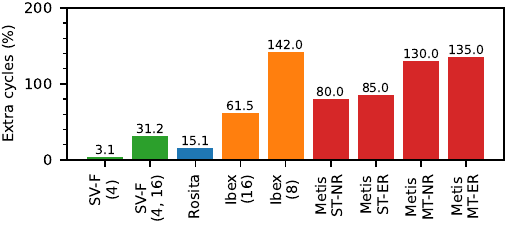}
    \caption{Execution time overhead on AES-128 encryption on ShuffleV compared with Rosita \cite{Shelton2021}, Secure Ibex \cite{ibex}, and Metis \cite{Antognazza2021} (lower is better). ``SV-F (4)''=ShuffleV-F (buffer size=4). ``SV-F (4,16)''=ShuffleV-F (buffer size=4, dummy interval=16). ``Ibex(16)''=Secure Ibex (dummy interval=16).}
    \label{figure:overhead_shufflev_vs_others}
\end{figure}

In all workloads, branch instructions are the primary cause of stalls, forcing the core to refill the shuffle buffer as described in Fig. \ref{figure:shufflev_wo_speculative_fetch}. Therefore, enabling branch prediction, such as with the `B' and/or `C' option, is likely to result in lower overhead, as shown in Fig. \ref{figure:overall_pt_cycle_overhead}. The \ouralg simulator and demo SoC currently support four branch prediction algorithms: always taken, always not taken, a static predictor based on branch offset, and a two-bit predictor. From our experiments, static branch prediction performs best for the CoreMark benchmark, achieving 82.00\% accuracy. For neural network and AES workloads, the two-bit branch predictor performs best, achieving 85.06\% and 71.27\% accuracy, respectively. The higher overhead observed in the AES-128 workload when the `B' option is enabled is likely due to low branch prediction accuracy, leading to significant checkpoint restore overhead (purple bar in Fig. \ref{figure:overall_pt_cycle_overhead}). Additionally, the high number of dependent memory access instructions in AES-128 prohibits checkpoint creation when branch prediction is enabled (orange bar in Fig. \ref{figure:overall_pt_cycle_overhead}). Despite these factors, \ouralg's overall overhead remains lower than existing work, as demonstrated in Fig. \ref{figure:overhead_shufflev_vs_others}. Note that determining the most accurate branch predictor for each specific workload is beyond the scope of this work, and interested readers are advised to refer to \cite{mittal2019survey}.

Fig. \ref{figure:overhead_shufflev_vs_others} compares the performance overhead of \ouralg on AES with SOTA works, including Secure Ibex with dummy instructions \cite{ibex}, Metis \cite{Antognazza2021}, the most efficient RISC-V core with code morphing engine, and Rosita \cite{Shelton2021}, an automatic code rewrite engine to protect the masked AES implementation. We choose \ouralg-F with buffer size equal to 4 for this comparison as it is the configuration used to perform security analysis in Sec. \ref{subsec:side_channel_eval}. When only shuffle is applied, the overhead of \ouralg is 3.1\%, 4.87x less than application specific protection (masked software implementation generated from Rosita) and 21.67x–47.29x less than generalized processors with built-in protection like Secure Ibex and Metis. 

When both shuffle and dummy instructions are applied, the overhead of \ouralg is 2.06x more than Rosita and 2.15x–4.7x less than generalized processors with built-in protection like Secure Ibex and Metis. Therefore, the findings indicate that the overhead of \ouralg is comparable to the application-specific defense like Rosita~\cite{Shelton2021}, while still being universal and user-friendly, eliminating the need for manual adjustments and software modifications.

To further demonstrate the scalability and compatibility of \ouralg on larger, real-world DNN architectures, Tab. \ref{tab:sfv_on_dnn_model} presents the execution time overhead (in cycles) when executing various models on \ouralg-F (bs=4) compared to the Ibex core. For all model architectures, we convert predefined models from Keras \cite{chollet2015keras} into TF Lite Micro format and set the input shape to 224x224x3, a common standard for RGB images in recent DNN literature. We include a range of models, from MobileNetV2 with 1.69M parameters (726 million CPU cycles) to ResNet50 with 25.61M parameters (23,045 million CPU cycles). Despite this significant variation in model size and computational complexity, the results indicate that \ouralg maintains a consistently low execution time overhead, ranging from 9.11\% to 13.15\% across all evaluated architectures. This highlights \ouralg's effectiveness in securing diverse and complex DNN workloads with minimal performance impact.

Unfortunately, it is not possible to compare the performance overhead of neural network workloads, due to the lack of prior related works on countermeasures against EM {SCAs} on neural networks at the software or microarchitecture level, with the exception of \cite{Brosch2022}, which also did not report performance numbers. We hope that the performance number presented in Fig. \ref{figure:overall_pt_cycle_overhead} and Tab. \ref{tab:sfv_on_dnn_model} could serve as a baseline for other future works on developing countermeasures against EM  {SCAs} on the neural network workload.

\begin{table}[]
\caption{Execution time overhead of \ouralg (bs=4) compared to the Ibex core on real-world DNN architecture.}
\label{tab:sfv_on_dnn_model}
\centering
\footnotesize
\begin{tabular}{lrrr}
\toprule
\multicolumn{1}{c}{\multirow{2}{*}[-2pt]{\textbf{Model Architecture}}} & \multicolumn{2}{c}{\textbf{Model Size}}    & \multicolumn{1}{c}{\multirow{2}{*}[-2pt]{\textbf{\makecell{\#Extra CPU\\Cycles}}}} \\ \cmidrule{2-3}
                   & \textbf{\#Params} & \textbf{\#CPU Cycles} &                                    \\
\midrule
MobileNetV2 ($\alpha$=0.35)    & 1.69M             & 726M          & 10.16\%\textsuperscript{$\pm$0.0018}                                      \\
MobileNetV2 ($\alpha$=0.75)    & 2.66M             & 1,978M        & 9.68\%\textsuperscript{$\pm$0.0018}                                        \\
EfficientNet B0    & 5.33M             & 4,175M        & 13.15\%\textsuperscript{$\pm$0.0010}                                        \\
NASNetMobile       & 5.33M             & 4,844M        & 9.11\%\textsuperscript{$\pm$0.0003}                                         \\
InceptionV3        & 23.85M            & 17,237M       & 9.64\%\textsuperscript{$\pm$0.0003}                                        \\
ResNet50           & 25.61M            & 23,045M       & 10.02\%\textsuperscript{$\pm$0.0004}                                        \\
\bottomrule
\end{tabular}
\end{table}

\begin{figure}
    \centering
    \includegraphics[width=0.95\linewidth]{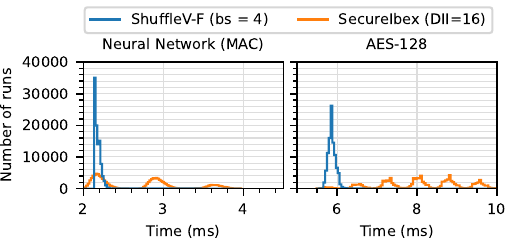}
    \caption{Distribution of execution time from 100,000 runs on \ouralg (bs=4) compared with Secure Ibex (DII=16) on MAC and AES-128 workloads.}
    \label{figure:runtime_variation}
\end{figure}

\textbf{Execution Time Variation. } While \ouralg's nondeterministic execution can introduce execution time variation due to shuffle buffer exhaustion and checkpoint restoring, the results in Tab. \ref{tab:sfv_on_dnn_model} demonstrate that this variation is negligible for large program segments. Furthermore, for shorter programs such as AES-128 and 5i5w MAC operations (from Sec. \ref{subsec:baseline_attack}), Fig. \ref{figure:runtime_variation} clearly illustrates a substantial reduction in execution time variation compared to Secure Ibex. These results demonstrate \ouralg's suitability for embedded and real-time systems that require predictable run-time behavior. Our instruction shuffling technique offers a distinct advantage over pure dummy instruction insertion, as it dynamically reorders actual program instructions rather than randomly inserting superfluous ones. This approach more efficiently utilizes the processor, leading to reduced execution time overhead and improved consistency in execution time.

\begin{table}
\caption{FPGA resource utilization of \ouralg compared to the Ibex \cite{ibex} and Secure Ibex \cite{ibex} core. BRAM utilization is 128 in all configurations. (``SV'' = ``\ouralg'', ``bs'' = ``buffer size'' and ``di'' = ``dummy instruction insertion interval'')}
\label{tab:fpga_resource_timing}
\centering
\footnotesize
\begin{tabularx}{\linewidth}{c|lcl}
\toprule
\textbf{Configuration}    & \multicolumn{1}{c}{\textbf{LUT as logic}} & \textbf{LUTRAM} & \multicolumn{1}{c}{\textbf{FF}} \\ 
\midrule
Ibex \cite{ibex}                   & 6547           & 96         & 5890    \\
Secure Ibex \cite{ibex}  & 6704 (+2.40\%)  & 96         & 5961 (+1.21\%)    \\
\midrule
SV (bs=2)                 & 7432 (+13.5\%)     & 140     & 6601 (+12.1\%)    \\ 
SV-MJB (bs=2)             & 8726 (+33.3\%)     & 140     & 6896 (+17.1\%)    \\
SV-F (bs=2)               & 7463 (+14.0\%)     & 140     & 6604 (+12.1\%)    \\
SV-FMJB (bs=2)            & 8599 (+31.3\%)     & 140     & 6897 (+17.1\%)    \\
SV-F (bs=2, di=16)        & 7598 (+16.1\%)     & 140     & 6695 (+13.7\%)    \\
\midrule
SV (bs=4)                 & 8032 (+22.7\%)     & 140     & 6910 (+17.3\%)    \\ 
SV-MJB (bs=4)             & 8888 (+35.8\%)     & 140     & 7215 (+22.5\%)    \\ 
SV-F (bs=4)               & 8079 (+23.4\%)     & 140     & 6915 (+17.4\%)    \\ 
SV-FMJB (bs=4)            & 8834 (+34.9\%)     & 140     & 7218 (+22.6\%)    \\
SV-F (bs=4, di=16)        & 8054 (+23.0\%)     & 140     & 7011 (+19.0\%)    \\
\midrule
SV (bs=8)                 & 8370 (+27.8\%)     & 140     & 7552 (+28.2\%)    \\ 
SV-MJB (bs=8)             & 9829 (+50.1\%)     & 140     & 7879 (+33.8\%)    \\
SV-F (bs=8)               & 8333 (+27.3\%)     & 140     & 7560 (+28.4\%)    \\ 
SV-FMJB (bs=8)            & 9705 (+48.2\%)     & 140     & 7886 (+33.9\%)    \\
SV-F (bs=8, di=16)        & 8586 (+31.1\%)     & 140     & 7655 (+30.0\%)    \\
\bottomrule
\end{tabularx}
\end{table}

\subsection{Hardware Resources and Timing} \label{subsec:hardware_resource}

To evaluate \ouralg hardware resource utilization, power consumption and side channel security, we developed \ouralg Demo SoC\footnote{Available at \url{https://github.com/nuntipat/ShuffleV-Demo-System}.} based on the Ibex demo system \cite{ibex_demo_soc}, a simple SoC design that combines the Ibex core with some basic peripheral to support device programming, debugging, and interfacing with external devices. 
Tab. \ref{tab:fpga_resource_timing} compares resource utilization of notable configurations of \ouralg Demo SoC against the Ibex Demo SoC. Both SoCs contain the same set of peripherals
, differing only in their CPU core.
The BRAM utilization is constant across all configurations as it depends solely on the size of the program memory. In all configurations, \ouralg runs at the same speed as the baseline Ibex core \cite{ibex} at 50 MHz. 
It's important to note that the percentage increase in resource utilization is relative to the overall SoC size. For example, Appendix \ref{appendix_fpga_resource} reveals that the \ouralg core constitutes only roughly half of the SoC's resources. Therefore, in real-world SoCs with more extensive peripherals and integrated memory, the core will occupy a much smaller proportion of the total area, making its area increase less significant overall.

\subsection{Power Consumption} \label{subsec:power_eval}

\begin{figure}
    \centering
    \includegraphics[width=0.9\linewidth]{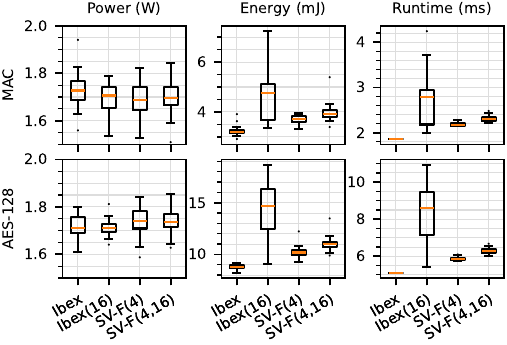}
    \caption{Power consumption, energy consumption, and execution time of \ouralg compared with Ibex \cite{ibex}, Secure Ibex (DII=16) (Ibex(16)) \cite{ibex}, ShuffleV-F (bs=4) (SV-F(4)), and ShuffleV-F (bs=4, DII=16) (SV-F(4, 16))}
    \label{figure:energy_comparison}
\end{figure}

We assess the effect of our design on power consumption by measuring the average power and energy consumption of a single neural network inference and AES encryption operation. 
Fig. \ref{figure:energy_comparison} shows the distribution of the power, energy per inference/encryption, and execution time obtained from 30 runs. In terms of power, both SecureIbex and \ouralg consume less average power than the unsecure Ibex. In the case of \ouralg, the reason is due to the additional stall cycles introduced by the shuffling logic, which reduce the average power. Similarly, Secure Ibex inserts a random number of dummy instructions that take different cycles to execute, resulting in lower average power consumption, e.g., a DIV instruction stalls the fetch and decode stages by up to 37 cycles but increases the energy consumption significantly (see Fig.\ref{figure:dummy_overhead} and Fig. \ref{figure:energy_comparison}).

In terms of energy consumption on the neural network workload, the Secure Ibex (Ibex(16)) consumes the highest energy, followed by \ouralg with dummy (SV-F(4,16)), \ouralg (SV-F(4)), and the Ibex (4.681mJ vs 3.962mJ vs 3.694mJ vs 3.242mJ). A similar trend can be observed on the AES workload, where the Secure Ibex consumes 14.53mJ, followed by SV-F(4,16) at 11.016mJ, SV-F(4) at 10.242mJ, and Ibex at 8.769mJ. In addition to its lower average energy consumption, \ouralg achieves a significantly lower variation in energy consumption due to lower execution time variations.

\subsection{Security Evaluation} \label{subsec:side_channel_eval}

We test the side-channel resistance of \ouralg from three perspectives: 1) Overall program shuffle; and 2) Critical operation shuffle; and 3) Resistance against correlation electromagnetic analysis attack (CEMA) using 1M traces.

\begin{figure}
    \centering
    \includegraphics[width=0.9\linewidth,trim={5 3 0 5},clip]{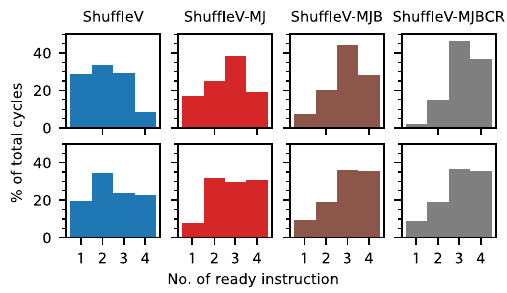}
    \caption{\update{Proportion of number of ready instructions in each processor cycle} while executing the ensemble model on TFLite Micro (top) and AES-128 encryption (bottom) on different configurations of \ouralg. \update{Results for other configuration are provided in Fig. \ref{figure:tfensem_fc_pt_cycle_vs_no_ready_inst}-\ref{figure:aes_pt_cycle_vs_no_ready_inst} in Appendix \ref{sec:shuffability_metrices}.}}
    \label{figure:summary_pt_cycle_vs_no_ready_inst}
\end{figure}

\textbf{Overall Program Shuffle.} 
We present the number of ready instructions (in $\%$) in each processor cycle in Fig. \ref{figure:summary_pt_cycle_vs_no_ready_inst}. The number of ready instructions is defined as the number of instructions that can be selected for execution in each clock cycle, which can range from 1 to $N$ (shuffle buffer's size). The higher number of ready instructions implies higher overall randomness.
When executing the neural network workload without applying any optimization (i.e., the vanilla \ouralg), 
28.66\% of cycle has 1 ready instruction and 8.45\% of cycle has 4 ready instruction, respectively. 
After enabling the M, J, and B options (Sec. \ref{subsec:summary_shufflev_configuration}), the percentage of cycles that have with only 1 ready instruction decreases to 7.69\%, while the percentage of cycles with 4 ready instructions increases to 28.01\%. These experimental results demonstrate the effectiveness of our optimization strategies in Sec. \ref{subsec:summary_shufflev_configuration} in improving overall randomness. Similar results are observed in the AES workload (Fig. \ref{figure:summary_pt_cycle_vs_no_ready_inst} (bottom)).

\begin{figure}
    \centering
    \includegraphics[width=0.9\linewidth,trim={0 0 0 0},clip]{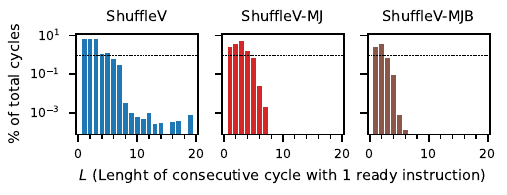}
    \caption{\update{Percent of total cycles with L consecutive cycles with 1 ready instruction while executing the ensemble model on TFLite Micro. Note that the sum of all bars in each plot is equal to the leftmost bar of the corresponding plot in Fig. \ref{figure:summary_pt_cycle_vs_no_ready_inst}.}}
    \label{figure:count_lenght_cycle_one_ready_half}
\end{figure}

To further validate the overall randomness, we propose the length of consecutive cycle with 1 ready instruction ($L$) as another metric to measure how long the core executes instruction without any shuffling. Fig. \ref{figure:count_lenght_cycle_one_ready_half} shows that it is extremely rare for the core to execute without shuffling when performing neural network inference (L\textgreater5 cycles for \textless1\% and L\textgreater7 cycles for \textless0.1\% of the total cycle). This experiment shows that while \ouralg may at time be unable to shuffle instructions due to true data dependency (Read-After-Write), a long sequence of non-shuffleable instructions is rare in practice, due to the nature of RISC ISA that requires additional instructions to load/store data from memory, increment the pointer address, etc. These instructions can be reordered to create run-time variation, even though the computation exhibits RAW dependency.

\begin{figure}
    \centering
    \includegraphics[width=0.9\linewidth]{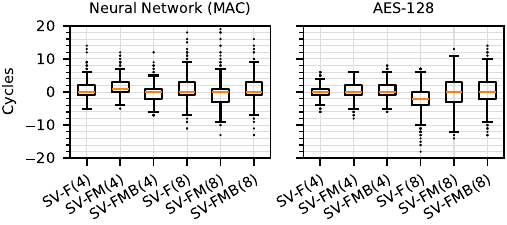}
    \caption{The amount of shuffling (in cycles) of the critical instructions in neural network inference (left) and AES-128 encryption (right) on different configurations of \ouralg.}
    \label{figure:sensitive_inst_move_num_cycle}
\end{figure}

\textbf{Critical Operation Shuffle.} We measure the amount of shuffling for critical instructions, i.e., these directly process secret values. To demonstrates how the overall randomness of the execution weaken the capability of attackers, we define the multiply-accumulate (MAC) operation of the neural network inference, as well as the load/store operation between the S-box and the state array as the critical operation. Using the shuffle buffer size equal to 4 and 8, Fig. \ref{figure:sensitive_inst_move_num_cycle} illustrates that the amount of shift for critical operations is approximately -13 to +19 cycles on the neural network workload, and -18 to +14 cycles on the AES-128 workload, depend on the \ouralg configuration and the overall program duration. These results demonstrate that, although \ouralg targets randomizing the overall program, it also affects the critical operations and can help adverse the ability of attack to conduct a successful attack.

\begin{table*}[]
\centering
\footnotesize
\caption{Results of performing a CEMA against AES-128 on unprotected Ibex \cite{ibex}, Secure Ibex \cite{ibex}, \ouralg-F with buffer size = 4 (SV-F (4)) and \ouralg-F with buffer size = 4 and dummy interval = 16 (SV-F (4,16)). The percentage after the configuration name indicates the overhead compared to the Ibex core. \xmark indicates an unsuccessful attack at 1M traces.}
\label{tab:aes_cema_summary_num_trace}
\resizebox{\linewidth}{!}{
\begin{tabular}{l|cccccccccccccccc}
\toprule
\textbf{\backslashbox{Config \xspace\quad}{Byte\#}} & \textbf{0} & \textbf{4} & \textbf{8} & \textbf{12} & \textbf{1} & \textbf{5} & \textbf{9} & \textbf{13} & \textbf{2} & \textbf{6} & \textbf{10} & \textbf{14} & \textbf{3} & \textbf{7} & \textbf{11} & \textbf{15} \\
\midrule
Ibex  & 300  & 300    & 400    & 150    & 400    & 200    & 100    & 100    & 150    & 250    & 250    & 250    & 1050   & 100    & 450    & 450  \\
Secure Ibex (61.5\%)  & 80k & 51k & 93k & 31k & 53k & 43k & 18k & 10k & 10k & 13k & 13k & 10k & 12k & 3k & 14k & 1k \\
SV-F (4) (3.1\%) (1\textsuperscript{st} run)  & 235k & \xmark & 385k   & \xmark & \xmark & \xmark & \xmark & \xmark & \xmark & \xmark & \xmark & \xmark & 170k   & 340k   & 310k  & \xmark \\
SV-F (4) (3.1\%) (2\textsuperscript{nd} run)  & \xmark & \xmark & \xmark & \xmark & \xmark & \xmark & \xmark & \xmark & 710k & \xmark & \xmark & \xmark & 360k   & \xmark & 100k & 530k \\
SV-F (4,16) (31.2\%) & 660k & \xmark & \xmark & \xmark & \xmark & \xmark & \xmark & \xmark & \xmark & \xmark & \xmark & \xmark & \xmark & \xmark & \xmark & \xmark \\
\bottomrule
\end{tabular}
}
\end{table*}

\begin{table}[]
\centering
\footnotesize
\caption{Results of performing a CEMA against MAC operation on unprotected Ibex \cite{ibex}, Secure Ibex \cite{ibex}, \ouralg-F (buffer size = 4) and \ouralg-F (buffer size = 4, dummy interval = 16). The percentage after the configuration name indicates the overhead compared to the Ibex core. \xmark indicates an unsuccessful attack at 1M traces. \textsuperscript{*1}The attack was performed on weights 1-2 simultaneously. \textsuperscript{*2}The attack yields five candidates with equal correlation (20\% success rate).}
\label{tab:dot_product_cema_summary_num_trace}
\begin{tabular}{l|ccccc}
\toprule
\textbf{\backslashbox{Config}{Weight\#}} & \textbf{1} & \textbf{2} & \textbf{3} & \textbf{4} & \textbf{5}\\
\midrule
Ibex  & 270  & 170  & 300 & 150 & 670  \\
Secure Ibex (37.5\%) & 37k\textsuperscript{*1} & 37k & 10k & 14k & 36k \\
SV-F (4) (13.7\%)    & 130k\textsuperscript{*1,2}& 130k\textsuperscript{*2} & 45k\textsuperscript{*2} & 100k\textsuperscript{*2} & \xmark     \\
SV-F (4,16) (41.8\%) & \xmark & \xmark & \xmark & \xmark & \xmark \\
\bottomrule
\end{tabular}
\end{table}

\begin{figure}[htpb]
\centering
    \includegraphics[width=0.87\textwidth]{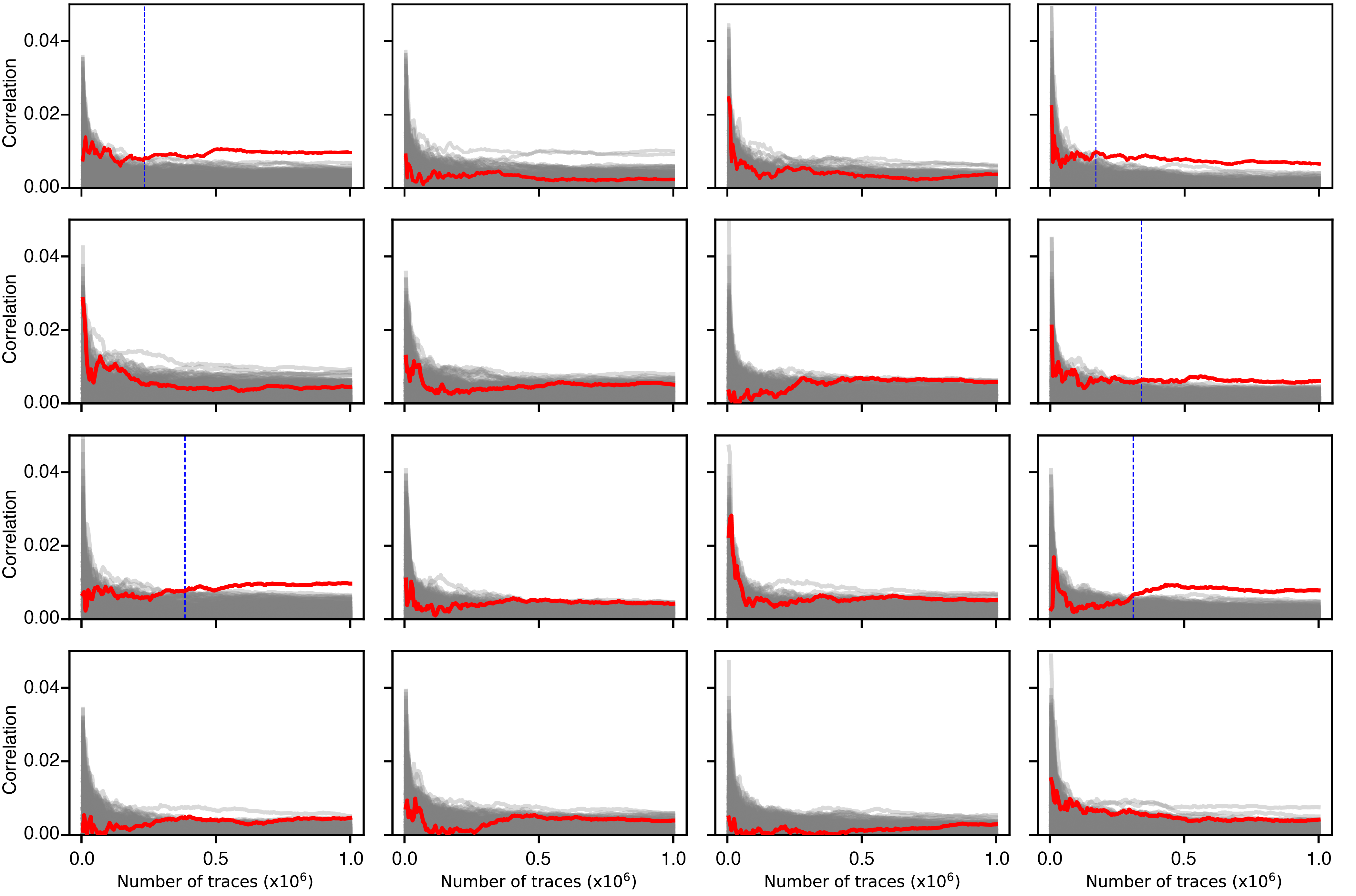}
    \caption{Result of performing a CEMA against AES on the \ouralg-F core (buffer size = 4). Blue dashed lines indicate the number of traces required for the successful attack. Otherwise, the attack is unsuccessful at 1M traces.}
    \label{figure:CEMA_aes_svf4}
\end{figure}

\begin{figure*}
    \includegraphics[width=\textwidth]{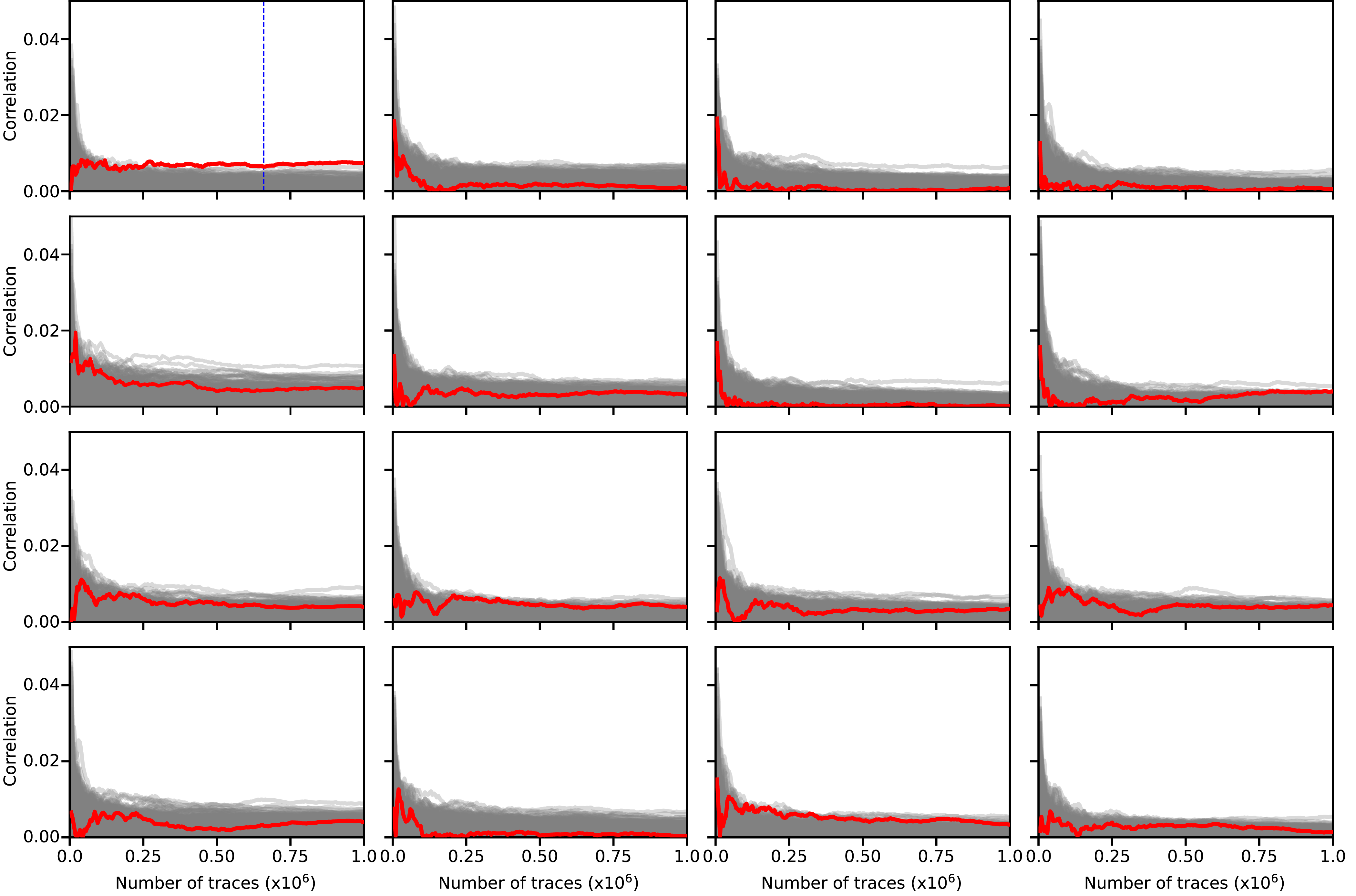}
    \caption{Result of performing a CEMA against AES on the \ouralg-F core (buffer size = 4, dummy interval = 16). Blue dashed lines indicate the number of traces required for the successful attack. Otherwise, the attack is unsuccessful at 1M traces.}
    \label{figure:CEMA_aes_svf4d16}
\end{figure*}

\textbf{Resistance against CEMA.} We validate the performance of \ouralg by conducting CEMA attack on the multiply-accumulate (MAC) operation and the SubBytes operation in the first round of AES-128 encryption. The measurements are conducted on the PYNQ-Z2 FPGA board running the \ouralg Demo SoC at 50MHz. We follow the experimental setup in Sec. \ref{subsec:general_experimental_setup}. Tab. \ref{tab:aes_cema_summary_num_trace} shows the CEMA results on the AES-128 encryption. From our experiment, the vanilla \ouralg (SV-F4) with 3.1\% performance overhead can protect approximately 12 out of 16 key bytes (Appendix. \ref{ap:cema_base} Fig. \ref{figure:CEMA_aes_svf4}). The number of traces required to perform the attack and the index of a byte in the key that can be successfully attacked change between subsequent runs, which proves that \ouralg execution is random. Note that even though several key bytes may leak in each run, it is still difficult for attackers to combine results from multiple runs to get the correct key as they do not know which byte is corrected. 
When considering \ouralg with dummy instructions, the performance overhead is 31.2\% and all except the first byte of the key can be successfully protected (Appendix. \ref{ap:cema_base} Fig. \ref{figure:CEMA_aes_svf4d16}). This is due to the limited shuffling space between the first byte and the trigger signal. Note that in this experiment, we assume a very strong attacker who can place the trigger signal precisely at the beginning of the encryption process, which is not feasible in practice. Thus, the actual attack success rate will be much lower. 

\begin{figure}
\centering
    \begin{subfigure}[b]{0.275\textwidth}
        \includegraphics[width=\textwidth]{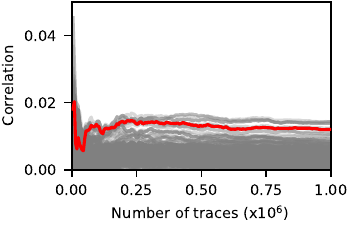}
    \end{subfigure}
    \begin{subfigure}[b]{0.275\textwidth}
        \includegraphics[width=\textwidth]{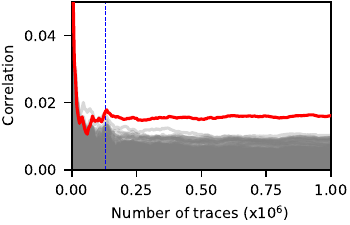}
    \end{subfigure}
    \begin{subfigure}[b]{0.275\textwidth}
        \includegraphics[width=\textwidth]{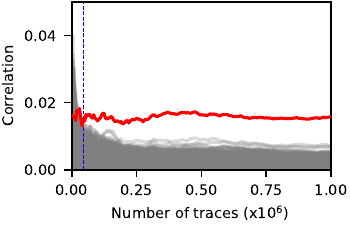}
    \end{subfigure}
    \begin{subfigure}[b]{0.275\textwidth}
        \includegraphics[width=\textwidth]{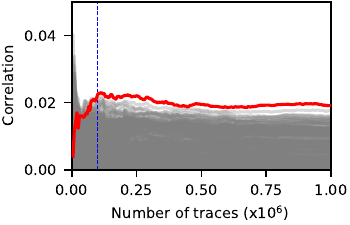}
    \end{subfigure}
    \begin{subfigure}[b]{0.275\textwidth}
        \includegraphics[width=\textwidth]{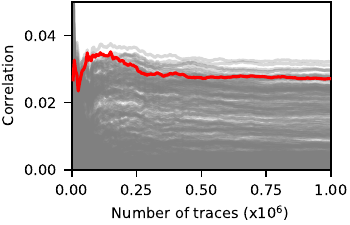}
    \end{subfigure}
    \caption{Result of performing a CEMA against 5i5w MAC on the \ouralg-F core (buffer size = 4). Blue dashed lines indicate the number of traces required for the successful attack. Otherwise, the attack is unsuccessful at 1M traces.} 
    \label{figure:CEMA_5i5w_svf4}
\end{figure}

\begin{figure*}
\centering
    \begin{subfigure}[b]{0.329\textwidth}
        \includegraphics[width=\textwidth]{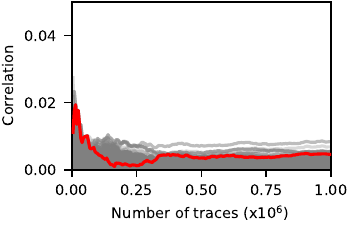}
    \end{subfigure}
    \begin{subfigure}[b]{0.329\textwidth}
        \includegraphics[width=\textwidth]{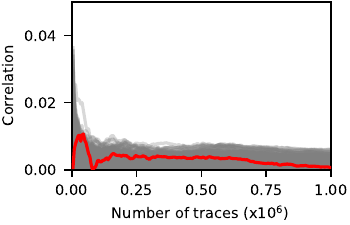}
    \end{subfigure}
    \begin{subfigure}[b]{0.329\textwidth}
        \includegraphics[width=\textwidth]{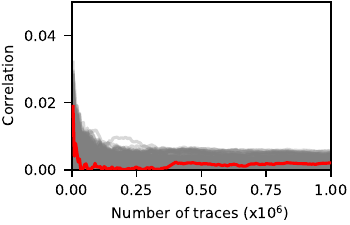}
    \end{subfigure}
    \begin{subfigure}[b]{0.329\textwidth}
        \includegraphics[width=\textwidth]{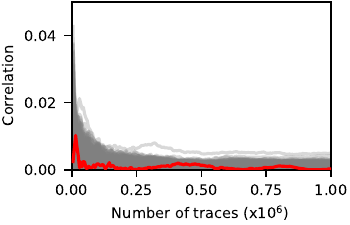}
    \end{subfigure}
    \begin{subfigure}[b]{0.329\textwidth}
        \includegraphics[width=\textwidth]{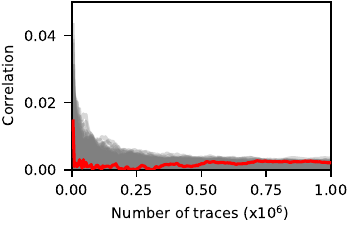}
    \end{subfigure}
    \caption{Result of performing a CEMA against 5i5w MAC on the \ouralg-F core (buffer size = 4, dummy interval = 16).} 
    \label{figure:CEMA_5i5w_svf4d16}
\end{figure*}

We show the CEMA results on the MAC in Tab. \ref{tab:dot_product_cema_summary_num_trace}. When considering \ouralg with a buffer size of 4 (i.e., SV-F (4)), the performance overhead is 13.7\%, and all weights from the fifth weight onwards can be successfully protected. The first weight can only be attacked in conjunction with an attack on weights 1 and 2. Attacking weights 2-4 yields five candidates with equal correlation values, resulting in a 20 percent success rate (Appendix. \ref{ap:cema_base} Fig. \ref{figure:CEMA_5i5w_svf4}). Again, the reason that weights 1–4 can be attacked is due to their close proximity to the trigger signal, and the success rate in reality will be lower.
For better defense, we test \ouralg with dummy instruction enabled (i.g., SV-F (4, 16)), which has a performance overhead of 41.8\% and could protect all weights (Appendix. \ref{ap:cema_base} Fig. \ref{figure:CEMA_5i5w_svf4d16}). Given that real-world neural networks have much more than 5 weights, the results show that both configurations can successfully protect neural network execution, and the choice of which configuration to use depends on the requirement and acceptable performance loss.

\section{Discussion and Future Work}

In this section, we discuss the limitations and potential misconceptions of \ouralg, proposing mitigation strategies and directions for future work.

\textbf{Compiler optimization:} A modern compiler may reorder instructions to maximize performance for target micro-architectures, e.g., the load instruction may be placed apart from the instructions that require the loaded value to prevent pipeline stall. In some cases, \ouralg may unintentionally void this optimization, resulting in performance loss. Further studies could be conducted to enhance the compatibility between \ouralg and compiler optimizations.

\textbf{Deterministic execution for embedded system:} Many embedded and real-time systems require predictable run-time behavior, i.e., low variation between runs. \ouralg improve upon SecureIbex significantly in this regard, as shown in Fig. \ref{figure:runtime_variation} and \ref{figure:energy_comparison} (right) by avoiding using long-running instructions, e.g., DIV as a dummy, and by adopting several strategies to prevent shuffle buffer stalls (see Sec. \ref{subsec:performance_opt}). If more control is needed, \ouralg allows developers to disable the protection through the Control and Status Registers when handling interrupts or performing time-critical operations (see Sec. \ref{subsec:hw_implementation}).

\textbf{\ouralg vs other side-channel attacks:} Although our threat model focuses on single-tenant embedded systems, which are not susceptible to software-based attacks like Spectre, we anticipate that \ouralg's non-deterministic execution and memory access pattern could extend its security benefits to mitigate timing and cache-based side-channel attacks. Applying \ouralg to desktop-class processors to validate this potential is a key direction for future research.

\update{\textbf{\ouralg vs Out-of-Order core:} %
\ouralg and traditional out-of-order and super-scalar cores are fundamentally different, since \ouralg reorders instructions randomly, whereas traditional out-of-order cores reorder instructions deterministically, i.e., they shuffle instructions in the same way across multiple runs, making them still vulnerable to CEMA attacks.}

\section{Conclusion}

This paper introduces \ouralg, a microarchitectural defense strategy against EM side-channel attacks (SCAs) in microprocessors. Employing the moving target defense (MTD) philosophy, \ouralg mitigates EM SCAs by randomly shuffling program instruction execution order and inserting dummy instructions. Unlike application-specific countermeasures, \ouralg offers automatic protection without algorithm modification or software recompilation. To accommodate diverse design needs, \ouralg provides various configurations enabling trade-offs between performance overhead and security. We developed an \ouralg simulator for rapid evaluation of different configurations, allowing users to analyze execution traces of any RISC-V binary. We implemented \ouralg on the open-source Ibex RISC-V core, enabling its use as a drop-in replacement in existing SoC designs. Furthermore, we developed an \ouralg Demo SoC and implemented it on a Xilinx XUP PYNQ-Z2 FPGA board to assess hardware resource footprint, performance, power consumption, and EM SCA resistance. Experimental results demonstrate \ouralg's successful protection of neural network model confidentiality and AES encryption keys with low overhead in performance, energy, and hardware resources.

\section*{Acknowledgements}
This work is supported in part by the U.S. National Science Foundation under Grants CNS-2153690, CNS-2247892, CNS-2239672, OAC-2319962, and CNS-2326597.

\bibliographystyle{IEEEtran}
\bibliography{refs}

\appendices

\newpage
\onecolumn
\section{CEMA Attack Results}\label{ap:cema_base}

\clearpage
\onecolumn

\section{Performance Overhead}\label{ap:performance_overhead}

\begin{figure}[h]
    \centering
    \includegraphics[width=\linewidth,trim={0 0 0 0},clip]{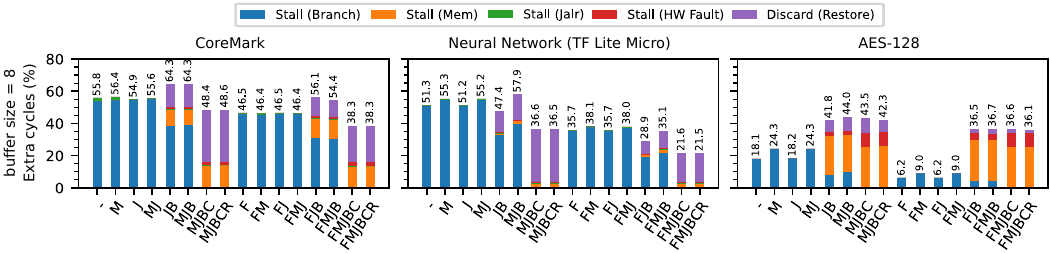}
    \caption{Performance overhead of different configuration of \ouralg on the CoreMark benchmark (left), neural network inference on TensorFlow Lite Micro (middle) and AES-128 encryption (right) with shuffle buffer size = 8}
    \label{figure:overall_pt_cycle_overhead_bs8}
\end{figure}

\section{Shuffablity metrices for all configurations} \label{sec:shuffability_metrices}

\begin{figure*}[htpb]
    \centering
    \includegraphics[width=0.85\linewidth,trim={0 0.31in 0 0},clip]{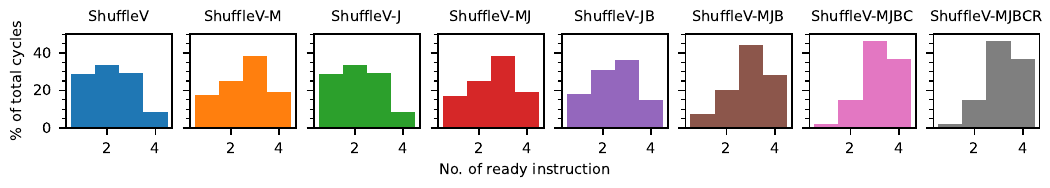}
    \includegraphics[width=0.85\linewidth]{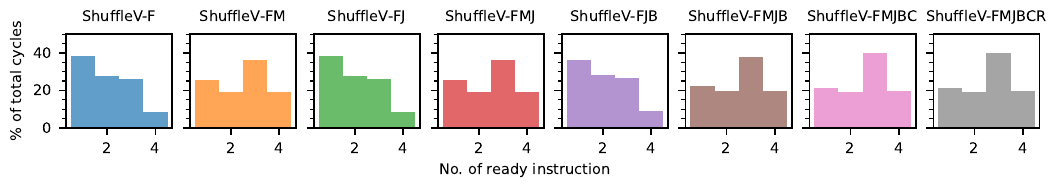}
    \caption{Proportion of CPU cycle with 1 - 4 ready instruction while executing the ensemble model on TFLite Micro}
    \label{figure:tfensem_fc_pt_cycle_vs_no_ready_inst}
\end{figure*}

\begin{figure*}[htpb]
    \centering
    \includegraphics[width=0.85\linewidth,trim={0 0.31in 0 0},clip]{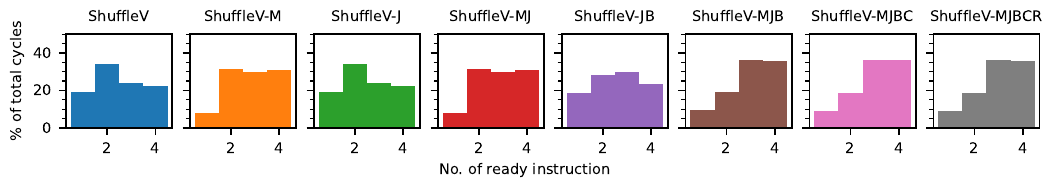}
    \includegraphics[width=0.85\linewidth]{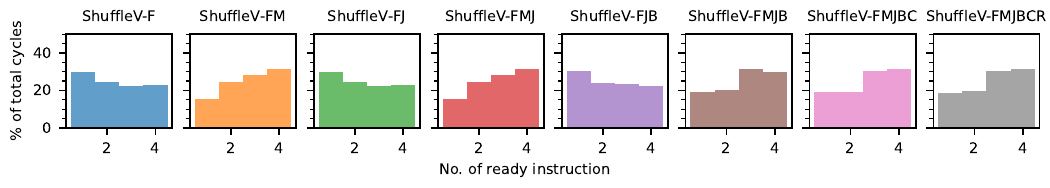}
    \caption{Proportion of CPU cycle with 1 - 4 ready instruction while executing the AES-128 encryption}
    \label{figure:aes_pt_cycle_vs_no_ready_inst}
\end{figure*}

\begin{figure*}[htpb]
    \centering
    \includegraphics[width=\linewidth]{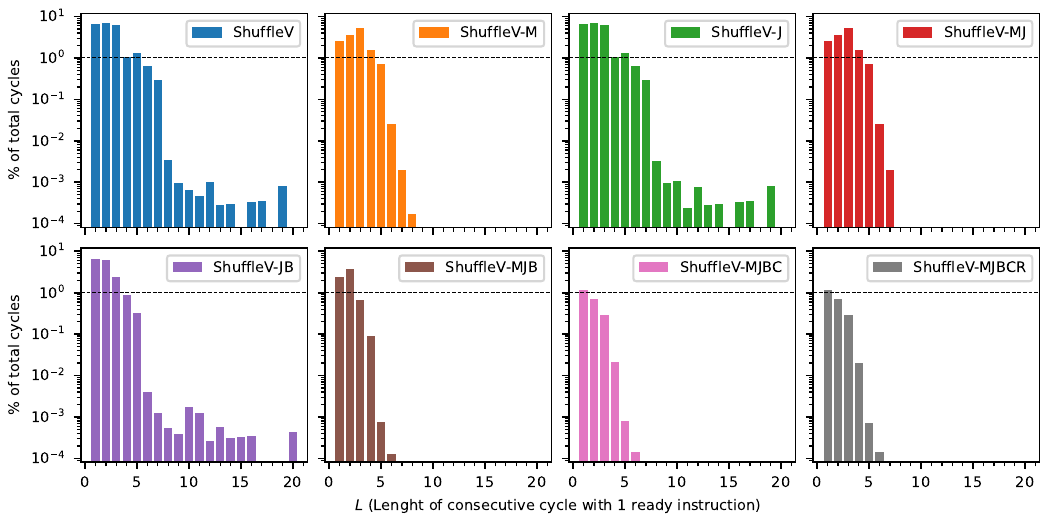}
    \caption{Percent of total cycles with L consecutive cycles with 1 ready instruction while executing the ensemble model on TFLite Micro. Note that the sum of all bars in each plot is equal to the leftmost bar of the corresponding plot in Fig. \ref{figure:tfensem_fc_pt_cycle_vs_no_ready_inst}.}
    \label{figure:tfensem_count_lenght_cycle_one_ready}
\end{figure*}

\begin{figure*}[htpb]
    \centering
    \includegraphics[width=\linewidth]{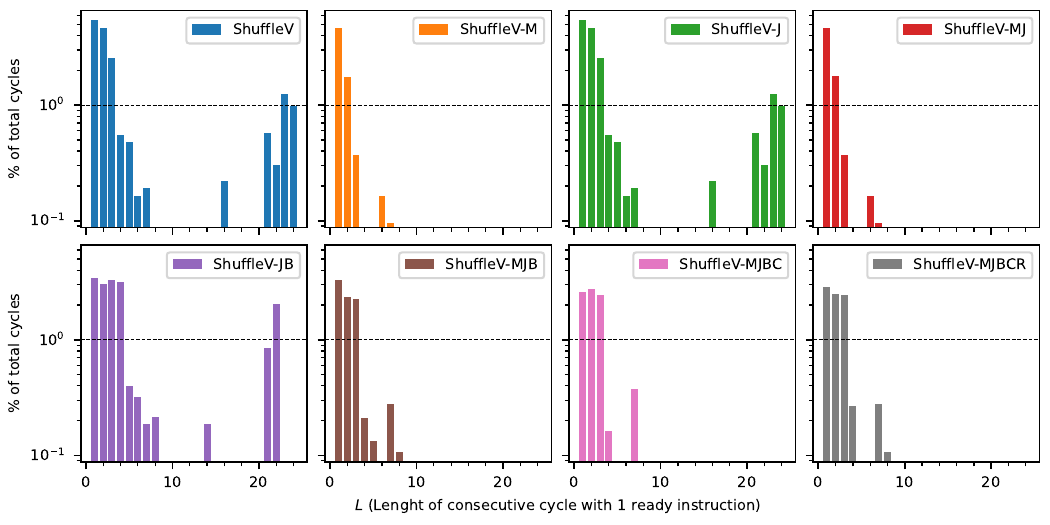}
    \caption{Percent of total cycles with L consecutive cycles with 1 ready instruction while executing the AES-128 encryption. Note that the sum of all bars in each plot is equal to the leftmost bar of the corresponding plot in Fig. \ref{figure:aes_pt_cycle_vs_no_ready_inst}.}
    \label{figure:aes_count_lenght_cycle_one_ready}
\end{figure*}

\clearpage
\onecolumn
\section{Ensemble model architecture} \label{appendix_model_arch}

\begin{figure}[h]
    \centering
    \includegraphics[width=0.7\linewidth,trim={0 12 0 0},clip]{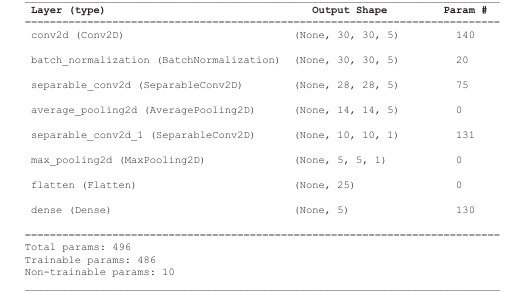}
    \caption{Architecture of the ensemble model used in our performance analysis.}
    \label{figure:tf_ensemble_model_arch}
\end{figure}

\section{FPGA resource utilization} \label{appendix_fpga_resource}

\begin{table}[h]
\centering
\caption{FPGA resource utilization of \ouralg compared to other RISC-V cores. \textsuperscript{*1}The Xilinx Zynq XC7020 FPGA combines an Arm Cortex-A9 processor with Artix-7 based programmable logic thus resource utilization from both FPGA devices should be comparable. \textsuperscript{*2}The BOOM result is assumed to be based on BOOMv1\cite{Celio:EECS-2015-167}. \textsuperscript{*3}The OPA and BOOM results are approximated from the figure presented in \cite{8977924}.}
\label{tab:hw_area_comp}
\begin{tabular}{l|l|l|l|r|r|r|c}
\toprule
\multicolumn{1}{c}{\textbf{Core}}      & \multicolumn{1}{|c}{\textbf{ISA}}      & \multicolumn{1}{|c}{\textbf{Type}}         & \multicolumn{1}{|c}{\textbf{FPGA Device\textsuperscript{*1}}}         & \multicolumn{1}{|c}{\textbf{LUTs}}         & \multicolumn{1}{|c}{\textbf{Registers}}     & \multicolumn{1}{|c}{\textbf{Reference}} \\ %
\midrule
PicoRV32  & RV32IM   & In-order     & Xilinx Artix XC7A35T & 1,765        & 1,075              & \cite{8760205}          \\ %
Ibex      & RV32IMC         & In-order     & Xilinx Zynq XC7Z020  & 3,161            &  1,933                   & -          \\
\textbf{\ouralg (bs=4)} & RV32IMC         & Out-of-order (Nondeterministic)     &  Xilinx Zynq XC7Z020                    & 4,411            & 2,993                     &   -        \\
RI5CY     & RV32IMC  & In-order     & Xilinx Artix XC7A35T & 6,748        & 2,577               & \cite{8760205}           \\ %
RSD       & RV32IM   & Out-of-order & Xilinx Zynq XC7Z020  & 15,379       & 8,584              & \cite{8977924}          \\ %
OPA\textsuperscript{*3}        & RV32IM   & Out-of-order & Xilinx Zynq XC7Z020  & 20,500      & 9,800             & \cite{8977924}           \\
BOOM\textsuperscript{*2}\textsuperscript{*3}     & RV32IMAC & Out-of-order & Xilinx Zynq XC7Z020  & 43,600       & 21,500             & \cite{8977924}           \\ %
\bottomrule
\end{tabular}
\end{table}

\end{document}